\documentclass[a4paper,10pt,aps,prd,twocolumn,superscriptaddress,nofootinbib,nobibnotes,longbibliography,floatfix]{revtex4-2}

\PassOptionsToPackage{dvipsnames}{xcolor}
\usepackage{orcidlink}
\usepackage{aasmacros}
\usepackage[utf8]{inputenc}
\usepackage[T1]{fontenc}

\usepackage{mathtools,amsmath,amssymb,amsfonts}
\usepackage{bm}
\usepackage{pifont}
\usepackage{lmodern}
\usepackage{mathrsfs}
\usepackage[normalem]{ulem}
\usepackage{lipsum}
\usepackage{subcaption}

\usepackage{xcolor}
\usepackage{graphicx} %
\usepackage{float}

\usepackage{hyperref}
\hypersetup{colorlinks=true, 
            citecolor=MidnightBlue,
            linkcolor=MidnightBlue, 
            urlcolor=MidnightBlue, 
            linktocpage=true}
\usepackage[capitalize]{cleveref}
\crefname{section}{Sec.}{Secs.}
\crefname{appendix}{App.}{Apps.}
\usepackage{array}
\usepackage{enumitem}
\usepackage[draft,defaultcolor=olive]{changes}

\usepackage{caption}
\usepackage{orcidlink}
\definechangesauthor{RB}

\begin{document}

\title{
Implications of the LISA stochastic signal from eccentric stellar mass black hole binaries in vacuum
}
\author{Ran Chen~\orcidlink{0000-0002-3370-8721}}
\email{ranchen@pmo.ac.cn}
\affiliation{Key Laboratory of Dark Matter and Space Astronomy, Purple Mountain Observatory, Chinese Academy of Sciences, Nanjing 210033, People's Republic of China}
\affiliation{School of Astronomy and Space Sciences, University of Science and Technology of China, Hefei 230026, People’s Republic of China}
\affiliation{SISSA, Via Bonomea 265, 34136 Trieste, Italy}

\author{Rohit~S.~Chandramouli~\orcidlink{0000-0001-5229-2752}}
\email{rchandra@sissa.it}
\affiliation{SISSA, Via Bonomea 265, 34136 Trieste, Italy}
\affiliation{INFN Sezione di Trieste, Trieste, Italy}
\affiliation{IFPU - Institute for Fundamental Physics of the Universe, Via Beirut 2, 34014 Trieste, Italy}
\author{Federico~Pozzoli~\orcidlink{0009-0009-6265-584X}}
\affiliation{Max Planck Institute for Gravitational Physics (Albert Einstein Institute), Am M\"uhlenberg 1, Potsdam 14476, Germany}

\author{Riccardo Buscicchio~\orcidlink{0000-0002-7387-6754}}
\affiliation{Dipartimento di Fisica ``G. Occhialini'', Università degli Studi di Milano-Bicocca, Piazza della Scienza 3, 20126 Milano, Italy}
\affiliation{INFN, Sezione di Milano-Bicocca, Piazza della Scienza 3, 20126 Milano, Italy}
\affiliation{Institute for Gravitational Wave Astronomy \& School of Physics and Astronomy, University of Birmingham, Birmingham, B15 2TT, UK}

\author{Enrico~Barausse~\orcidlink{0000-0001-6499-6263}}
\affiliation{SISSA, Via Bonomea 265, 34136 Trieste, Italy}
\affiliation{INFN Sezione di Trieste, Trieste, Italy}
\affiliation{IFPU - Institute for Fundamental Physics of the Universe, Via Beirut 2, 34014 Trieste, Italy}

\date{\today}%

\begin{abstract}
Astrophysical formation channels of stellar-mass binary black holes (sBBHs) can induce significant orbital eccentricities in their early inspiral. We analyze the implications on the stochastic gravitational-wave background (SGWB) from unresolved sBBHs, which can be detected with the Laser Interferometer Space Antenna (LISA).
We develop an improved SGWB model for the case of an idealized Dirac-delta eccentricity distribution, and extend it to the more astrophysical case of a thermal distribution. 
Using a fully Bayesian framework, we find that, if all binaries have a high initial eccentricity $e_0 \gtrsim 0.9$ at an orbital frequency of $f_{\rm orb} = 10^{-4}\,\mathrm{Hz}$, the resulting SGWB can be robustly distinguished from a background of quasi-circular sBBHs.
For a thermal eccentricity distribution, the SGWB is consistent with a circular model when binaries form at $f_{\rm orb} = 10^{-5}\,\mathrm{Hz}$, but leads to significant systematic biases if formation occurs at $f_{\rm orb} = 10^{-4}\,\mathrm{Hz}$. 
We also show that, when eccentricity is properly accounted for, environmental effects such as dynamical friction can be distinguished from vacuum evolution, but only for sufficiently dense environments with gas densities $\rho \gtrsim 10^{-7}\,\mathrm{g\,cm^{-3}}$.
Finally, we show that a LISA detection of the sBBH SGWB would place an upper bound on the maximum eccentricity of the sBBH population in the band of ground-based detectors, with direct implications for template modeling and data analysis.
Our results highlight the importance of incorporating eccentricity in SGWB modeling to enable accurate astrophysical interpretation of LISA observations.
\end{abstract}

\maketitle

\section{Introduction}

The detection of gravitational waves (GWs) from compact binary coalescences by the LIGO-Virgo-KAGRA (LVK) collaboration~\cite{LIGOScientific:2016aoc,LIGOScientific:2020ibl,LIGOScientific:2018mvr,KAGRA:2021vkt,KAGRA:2021duu} has opened a new window onto the population of stellar-mass binary black holes (sBBHs).
With the recent GWTC-4.0 catalog~\cite{LIGOScientific:2025slb} and updated population analysis~\cite{LIGOScientific:2025pvj} bringing the total number of confident detections above 200, a picture of the sBBH population is emerging with increasing clarity.
Multiple formation channels contribute to this population: isolated binary evolution through common-envelope phases~\cite{Belczynski:2016obo,Mandel:2018hfr}, dynamical encounters in dense stellar environments such as globular clusters~\cite{Rodriguez:2016kxx,Samsing:2017xmd,Zevin:2018kzq}, Kozai-Lidov oscillations in hierarchical triples~\cite{Naoz:2012bx,Antonini:2017ash}, and mergers facilitated by the gaseous disks of active galactic nuclei (AGNs)~\cite{Levin:2006uc,Bartos:2016dgn,Tagawa:2019osr}.
Crucially, these channels predict different orbital properties for the merging binaries, and in particular different distributions of orbital eccentricity at formation~\cite{Samsing:2017xmd,Zevin:2021rtf}.

The Laser Interferometer Space Antenna (LISA)~\cite{amaro2017laser}, a planned space-based gravitational-wave (GW) observatory operating in the mHz band, will observe sBBHs years to decades before they merge in the LVK band~\cite{Sesana:2016ljz}.
At these early stages of the inspiral, astrophysical environments can significantly affect the orbital evolution through dynamical friction, hydrodynamic drag, and gas accretion~\cite{Barausse:2007dy,Barausse:2014tra}.
These effects have been shown to leave detectable imprints on the GW signals from individually resolvable sBBHs in the LISA band~\cite{Toubiana:2020drf,Caputo:2020irr,Sberna:2022qbn}.
However, the vast majority of sBBHs will not be individually resolvable, instead contributing to a stochastic gravitational-wave background (SGWB)~\cite{Allen:1997ad,Phinney:2001di,Maggiore:2007ulw}.
Recent studies have shown that this background is detectable by LISA, with the vacuum amplitude measurable at percent-level precision~\cite{Babak:2023lro,Lehoucq:2023zlt,Pieroni:2020rob}.
Moreover, in our previous paper~\cite{Chen:2025qyj}, we showed that environmental effects -- dynamical friction and gas accretion -- can also imprint detectable deviations in the spectral shape of the SGWB from quasi-circular sBBHs.
In particular, dynamical friction from gas densities $\rho \gtrsim \mbox{a few}\, \times 10^{-10}\,\mathrm{g\,cm^{-3}}$, typical of AGN accretion disks, was found to be measurable, while the effect of gas accretion was shown to be undetectable for astrophysically motivated Eddington ratios.

However, the analysis in Ref.~\cite{Chen:2025qyj} assumed that all sBBHs evolve on quasi-circular orbits.
In general, sBBH formation channels can induce significant orbital eccentricities during the early inspiral~\cite{Samsing:2017xmd,Zevin:2021rtf} (see also Refs.~\cite{Bonetti:2017dan,Bonetti:2017lnj,Bonetti:2018tpf} for the massive black hole case).
While eccentricity is radiated away efficiently by GW emission, so that binaries are effectively circularized by the time they reach the LVK band~\cite{Peters:1963ux}, in the LISA mHz band the residual eccentricity can still be non-negligible.
The effect of eccentricity on the SGWB has been studied extensively in the literature, particularly for supermassive black hole binaries in the pulsar timing array (PTA) band~\cite{Enoki:2006kj,Chen:2016zyo} and for extreme mass ratio inspirals~\cite{Bonetti:2020jku}.
For sBBHs in the LISA band, Ref.~\cite{Liang:2025wfj} recently studied the detection of eccentric SGWBs using eccentricity distributions from different channels.
Eccentric binaries emit GW power across multiple harmonics of the orbital frequency, redistributing energy from the dominant quadrupole into higher harmonics~\cite{Peters:1963ux,Enoki:2006kj}.
This redistribution suppresses the SGWB at low frequencies relative to the quasi-circular case, producing a characteristic spectral turnover whose frequency depends on the initial eccentricity and formation orbital frequency.

This low-frequency suppression raises a critical point: both eccentricity and environmental effects such as dynamical friction suppress the SGWB at low frequencies, though through physically distinct mechanisms.
Eccentricity redistributes GW power among harmonics, while environmental effects open additional energy-loss channels that deplete the GW flux.
Nevertheless, their spectral signatures can appear qualitatively similar, potentially leading to degeneracies in the interpretation of the observed SGWB.
This degeneracy was identified as an important open question in Ref.~\cite{Chen:2025qyj}, and constitutes one of the main motivations for the present work.

In this paper, we develop improved models for the SGWB from eccentric sBBHs and perform a comprehensive Bayesian analysis to quantify the measurability of eccentricity with LISA and its degeneracy with environmental effects.
Our main contributions are as follows.
(i) We construct an improved phenomenological fitting model for the eccentric SGWB spectrum, valid for all eccentricities from the quasi-circular limit to $e_0 \sim 1$, which improves upon previous fits~\cite{Chen:2016zyo} by correctly capturing the asymptotic behavior at both low and high frequencies.
(ii) We extend the analysis from an idealized Dirac-delta eccentricity distribution to the more astrophysically motivated thermal distribution~\cite{Jeans1919,Ambartsumian1937,Heggie1975}, expected from dynamical formation channels.
(iii) Using a fully Bayesian framework, we find that if all binaries share a high initial eccentricity $e_0 \gtrsim 0.9$ at an orbital frequency of $f_{\rm orb} = 10^{-4}\,\mathrm{Hz}$, the resulting SGWB is robustly distinguishable from a quasi-circular background.
For a thermal eccentricity distribution, the SGWB is consistent with a circular model when binaries form at $f_{\rm orb} = 10^{-5}\,\mathrm{Hz}$, but leads to significant systematic biases in the inferred vacuum parameters if formation occurs at $f_{\rm orb} = 10^{-4}\,\mathrm{Hz}$.
(iv) We investigate the degeneracy between eccentricity and environmental effects by injecting SGWB signals containing dynamical friction and recovering with an eccentric vacuum model.
We show that when eccentricity is properly accounted for, dynamical friction can be distinguished from vacuum evolution, but only for sufficiently dense environments with $\rho \gtrsim 10^{-7}\,\mathrm{g\,cm^{-3}}$.
(v) We further show that a quasi-circular sBBH SGWB detected by LISA would translate into an upper bound on the population eccentricity at ground-based detector frequencies, excluding a uniform eccentricity distribution with $e^{\max}_{20\,{\rm Hz}}\gtrsim 10^{-2}$.
This paper is organized as follows.
In \cref{subsec:overview}, we review the computation of the SGWB from eccentric binaries, and in \cref{subsec:modeling_ecc_SGWB,subsec:modeling_ecc_SGWB_thermal} we develop our improved SGWB models for the Dirac-delta and thermal eccentricity distributions.
In \cref{sec:DA}, we present the data analysis framework and our Bayesian parameter estimation results, including the measurability of eccentricity, the biases from quasi-circular mismodeling, and the degeneracies with environmental effects.
In \cref{sec:erefmax}, we show how LISA observations of the sBBH SGWB can place an upper limit on the maximum eccentricity of the sBBH population at ground-based detector frequencies.
We present our conclusions in \cref{sec:conclusion}.

Throughout this paper, we adopt geometric units with $G=1=c$, and the conventions $m=m_1+m_2$ for the total mass, $q = m_2/m_1 < 1$ for the mass ratio, $\eta = m_1 m_2/m^2$ for the symmetric mass ratio, $\mathcal{M} = m \eta^{3/5}$ for the chirp mass, with the masses always expressed in the source's rest frame.

\section{SGWB of Eccentric binaries}
Computing the SGWB from eccentric binaries has been studied extensively in the literature.
In Sec.~\ref{subsec:overview}, we briefly review the basics of the relevant calculations, following Refs.~\cite{Phinney:2001di,Enoki:2006kj,Huerta:2015pva,Chen:2016zyo}, establishing the definitions and notations used in this work.
Following that, in Sec.~\ref{subsec:modeling_ecc_SGWB}, we discuss more details of the eccentric SGWB model that we develop, which improves on previous works.
\subsection{Overview}\label{subsec:overview}
Assuming a stochastic gravitational-wave background that is isotropic, unpolarized, and stationary, the present-day energy density per logarithmic frequency interval can be expressed in terms of the dimensionless energy density $\Omega_{\mathrm{GW}}(f)$, defined as ~\cite{Phinney:2001di}
\begin{equation}
\Omega_{\mathrm{GW}}(f) \equiv \frac{1}{\rho_c} \frac{d \rho_{\mathrm{GW}}}{d \log f},
\end{equation}
where $\rho_c=3 H_0^2 /(8 \pi)$ is today's Universe critical energy density.

By expressing the total gravitational-wave energy density in terms of the GW amplitude, one can define the dimensionless characteristic strain $h_c(f)$, which quantifies the amplitude per unit logarithmic frequency interval.
The corresponding relation between $\Omega_{\mathrm{GW}}(f)$ and $h_c(f)$ is given by~\cite{Maggiore:1999vm}
\begin{equation} \label{eqn:Omega_hc}
\Omega_{\mathrm{GW}}(f)=\frac{2 \pi^2}{3 H_0^2} f^2 h_c^2(f).
\end{equation}

As shown in Ref.~\cite{Phinney:2001di}, the present-day energy density must be equal to the sum of energy densities radiated at each redshift
$z$, weighted by $1/(1 + z)$ to account for cosmological redshift.
Accordingly, the characteristic strain can be expressed as
\begin{equation}
\label{hc_basic}
h_c^2(f)=\frac{4}{\pi f^2}\left.\int_0^{\infty} \frac{dn}{dz} \frac{1}{1+z}\left(f_r \frac{d E_{\mathrm{GW}}}{d f_r}\right)\right|_{f_r} d z,
\end{equation}
where $f_r$ is the GW frequency in the rest frame of the source, and $dn/dz$ is the comoving number density per unit redshift.
In the rest frame of a binary with masses $m_1$, $m_2$, orbital eccentricity $e$, and orbital frequency $f_{\mathrm{orb}}$, the total power radiated by GW emission to leading post-Newtonian (PN) order is given by~\cite{Peters:1963ux,Maggiore:1999vm}
\begin{equation}
P_{\rm GW} =\frac{32}{5}(2\pi f_{\mathrm{orb}}\mathcal{M})^{10 / 3}F(e),
\end{equation}
where $\mathcal{M}$ is the binary's chirp mass with the eccentricity induced enhancement factor $F(e)$ given by
\begin{equation}
\begin{aligned}
F(e)=\frac{1}{\left(1-e^2\right)^{7 / 2}}\left(1+\frac{73}{24} e^2+\frac{37}{96} e^4\right).
\end{aligned}
\end{equation}

With the help of Kepler's second law, the evolution of the orbital frequency $f_{\mathrm{orb}}$, in the rest frame, to leading PN order is given by
\begin{equation}
\label{eqn:dforbdt}
\frac{df_{\mathrm{orb}}}{dt}=\frac{96}{5} \frac{F(e)}{2 \pi} \mathcal{M}^{5 / 3} (2\pi f_{\mathrm{orb}})^{11 / 3}.
\end{equation}
For a given initial eccentricity $e_{0}$ at a fiducial initial orbital frequency $f_{\mathrm{orb},0}$, the evolutions of the orbital
frequency $f_{\mathrm{orb}}$ and the eccentricity $e$ are related through~\cite{Peters:1963ux}
\begin{equation}
\label{eqn:e_evolution}
\dfrac{f_{\mathrm{orb}}}{f_{\mathrm{orb},0}}=\left\{\dfrac{1-e_0^2}{1-e^2}\left(\dfrac{e}{e_0}\right)^{\frac{12}{19}}\left[\dfrac{1+\frac{121}{304} e^2}{1+\frac{121}{304} e_0^2}\right]^{\frac{870}{2299}}\right\}^{-3 / 2}.
\end{equation}

For an eccentric binary, the emitted GW is no longer instantaneously monochromatic but is instead distributed among integer harmonics, $f_{r}=jf_{\mathrm{orb}}$.
Following~\cite{Enoki:2006kj,Maggiore:2018sht,Huerta:2015pva,Chen:2016zyo}, from~\cref{eqn:e_evolution}, one can then determine the corresponding eccentricity such that the $j$-th harmonic (in source frame) coincides with the observed GW frequency $f = f_r / (1+z)$, which is a restatement of the stationary-phase-approximation for eccentric inspirals~\cite{Yunes:2009yz}, translating to,
\begin{align} \label{eqn:eccn}
 e_{j} &\equiv e_{j}\left( \dfrac{f_{\mathrm{orb}} }{ f_{\mathrm{orb},0}}, e_0\right) =e_{j}\left( \dfrac{f_r}{ j f_{\mathrm{orb},0}}, e_0\right) \\ \nonumber
 &= e_{j}\left( \dfrac{f(1+z) }{ j f_{\mathrm{orb},0}}, e_0\right),   
\end{align}
where $e_j$ is determined by inverting~\cref{eqn:e_evolution}.
The total power can be expressed as a sum over individual harmonics, with the power emitted in the $j$-th harmonic given by~\cite{Maggiore:1999vm,Peters:1963ux}
\begin{equation}
\label{eqn:npower}
P_{\mathrm{GW},j} \equiv \dfrac{dE_{\mathrm{GW},j}}{dt} =\frac{32}{5}(2\pi f_{\mathrm{orb}}\mathcal{M} )^{10 / 3} g(j, e_j), 
\end{equation}
where 
\begin{equation}
\begin{aligned}
g(j, e) & =\frac{j^4}{32}\left[\left\{J_{j-2}(j e)-2 e J_{j-1}(j e)\right.\right. \\
& \left.+\frac{2}{j} J_j(j e)+2 e J_{j+1}(j e)-J_{j+2}(j e)\right\}^2 \\
& +\left(1-e^2\right)\left\{J_{j-2}(j e)-2 J_j(j e)+J_{j+2}(j e)\right\}^2 \\
& \left.+\frac{4}{3 j^2} J_j^2(j e)\right].
\end{aligned}
\end{equation}

Thus, the energy spectrum results from summing over the contributions of all harmonics~\cite{Enoki:2006kj,Huerta:2015pva,Chen:2016zyo}, i.e.
\begin{equation}
\begin{aligned}
\frac{d E_{\mathrm{GW}}}{d f_{r}} &= \sum_{j=1}^{\infty}\frac{dE_{\mathrm{GW},j}}{dt}\frac{dt}{df_{\mathrm{orb}}}\frac{df_{\mathrm{orb}}}{df_r}. \label{eqn:dEdf_sum_ecc}
\end{aligned}
\end{equation}
From~\cref{eqn:dforbdt,eqn:npower} and the relation between $f_r$, $f_{\mathrm{orb}}$, and $f$, the energy spectrum then becomes~\cite{Enoki:2006kj,Huerta:2015pva,Chen:2016zyo} (see Appendix~\ref{app:validity_reaction} for an alternative derivation)
\begin{equation}
\begin{aligned}
\label{eqn:energy_spectrum}
\frac{d E_{\mathrm{GW}}}{d f_{r}} &= \frac{\mathcal{M}^{5/3} \pi^{2/3}}{3f^{1/3}(1+z)^{1/3}} S(f,f_{\mathrm{orb},0},e_0,z),\\
\end{aligned}
\end{equation}
where the eccentricity dependence is captured by the function $S \equiv S(f,f_{\mathrm{orb},0},e_0,z) $ given by
\begin{align}
\label{eqn:S_sum}
S(f,f_{\mathrm{orb},0},e_0,z) = \sum_{j=1}^{\infty}\frac{g(j,e_j)}{F(e_j)(j/2)^{2/3}}.
\end{align}
In the circular limit $e_0 \rightarrow 0$, we have that $S \rightarrow 1$, which results in the familiar expression~\cite{Phinney:2001di} for the energy spectrum $dE_{\rm GW}/df_r = \mathcal{M}^{5/3} \pi^{2/3} / (3 f^{1/3} (1+z)^{1/3})$.
Substituting the expression for the energy spectrum from~\cref{eqn:energy_spectrum} into~\cref{hc_basic}, and using the comoving differential number density of merging BBHs per unit redshift and unit chirp mass $d^2 n / (dz d\mathcal{M})$, we have 
\begin{equation}
\begin{aligned}
\label{eqn:hc_dndzdMc}
h_c^2(f)
=\frac{4}{3\pi^{1/3}f^{4/3}}\int dzd\mathcal{M}\frac{d^2n}{dzd\mathcal{M}} \frac{\mathcal{M}^{5/3}S}{(1+z)^{1/3}}.
\end{aligned}
\end{equation}
Given~\cref{eqn:hc_dndzdMc}, one can easily extend the result to an arbitrary eccentricity distribution by integrating over $d^3n/(dz d \mathcal{M} d e_0)$, which gives
\begin{equation}
\begin{aligned}
h_c^2(f)= & \frac{4}{3\pi^{1/3}f^{4/3}}\int d z d \mathcal{M} de_{0} \frac{d^3 n}{d z d \mathcal{M}de_{0}}  \frac{\mathcal{M}^{5/3}S}{(1+z)^{1/3}}.
\label{eqn:hc_dndzdMcde0}
\end{aligned}
\end{equation}
This completes the derivation of a SGWB from eccentric inspiralling binaries.

We see that the contribution to the energy spectrum at an observed frequency $f$ from a single binary is effectively generated by summing over all contributions of the same binary but with its eccentricity replaced by $e_j$ at a corresponding orbital frequency $f_{\rm orb} = f_r / j$.
Note that the waveform of a single binary can exhibit interference between its harmonics due to radiation reaction.
However, our expressions do not contain such interference from cross terms between the harmonics in the Fourier expansion.
This is because for an isotropic stochastic background, one has to average over all possible source orientations, which causes these cross terms to vanish.
Starting from Phinney's theorem~\cite{Phinney:2001di}, we prove this in Appendix~\ref{app:validity_reaction} and therein also present an alternative derivation of the energy spectrum given by~\cref{eqn:energy_spectrum}.
Further, in Appendix~\ref{app:validity_PN}, we also verify that the leading PN description for the energy spectrum is sufficiently accurate for our work.

\subsection{Modeling of eccentric SGWB: Dirac-delta distribution for eccentricity} \label{subsec:modeling_ecc_SGWB}
We first focus on the case where the eccentricity follows a Dirac-delta distribution, such that all sources have the same eccentricity $e_0$ at a given orbital frequency $f_{\rm orb,0}$.
Although this is idealized, it will help illustrate the phenomenology of how eccentricity influences the SGWB. 
Further, such simplification provides a useful starting point to incorporate generic eccentricity distributions and develop templates to search for eccentricity effects in the LISA stochastic signal from sBBHs.

With the eccentricity following a Dirac-delta distribution, we can use~\cref{eqn:hc_dndzdMc} to compute $h_c^2(f)$.
One can in principle perform this by brute force: draw samples $(\mathcal{M},z)$ from the distribution $d^2n /(dz d\mathcal{M})$, evaluate the sum for $S(f,f_{\rm orb}, e_0,z)$ in~\cref{eqn:S_sum} for an optimal $j_{\rm max}$, and approximate the integral in~\cref{eqn:hc_dndzdMc} with a Monte Carlo sum.
However, as noted by previous works, in particular in Refs.~\cite{Huerta:2015pva,Chen:2016zyo}, the function $S(f,f_{\rm orb}, e_0,z)$ has a simple scaling behavior for different values of $e_0$ and $f_{\rm orb,0}$.
This scaling can also be seen in $h^2_c(f)$ by formally a considering a Dirac-delta distribution for the sources with $d^2 n/(d z d \mathcal{M})=\delta(\mathcal{M}-\mathcal{M}_0) \delta(z-z_0)/\mathrm{Mpc}^3$, which results in
\begin{equation}
\begin{aligned}
\label{eqn:hc_delta}
h_c^2(f)
=h^2_{c,\mathrm{cir}} \times S(f,f_{\mathrm{orb},0},e_0,z_0),
\end{aligned}
\end{equation}
where
\begin{equation}\label{eqn:hc_circ}
\begin{aligned}
h^2_{c,\mathrm{cir}} = \frac{4}{3\pi^{1/3}f^{4/3}\mathrm{Mpc}^{3}} \frac{\mathcal{M}_0^{5/3}}{(1+z_0)^{1/3}}.
\end{aligned}
\end{equation}
We first compute an expensive computation for a reference $h_{c,\rm ref}^2(f)$ by evaluating~\cref{eqn:hc_delta} with an optimal $j_{\max}$.
Following this, we fit $h_{c,\rm ref}^2(f)$ with a suitable ansatz and suitably rescale to compute $h^2_c(f)$ for other values of $e_0$ and $f_{\rm orb,0}$.
We follow Ref.~\cite{Chen:2016zyo} for the rescaling:
\begin{equation}
\begin{aligned}
\label{eqn:hc_general}
h_c^{2}(f)
=h^{2}_{c,\mathrm{ref}}(f \frac{f_{p,\mathrm{ref}}}{f_{p}})\left( \frac{f_{p}}{f_{p,\mathrm{ref}}}\right)^{-4/3},
\end{aligned}
\end{equation}
where $f_{p} \equiv f_p (f_{\mathrm{orb},0},e_0)$ denotes the peak frequency of $S(f,f_{\mathrm{orb},0},e_0,z)$ as described in Ref.~\cite{Huerta:2015pva}, given by
\begin{equation}\label{eqn:fpeak}
\frac{f_p}{f_{\mathrm{orb},0}}=\frac{1293}{181}\left[\frac{e_0^{12 / 19}}{1-e_0^2}\left(1+\frac{121 e_0^2}{304}\right)^{\frac{870}{2299}}\right]^{3 / 2}.
\end{equation}

To ensure that the scaling is robust across all possible $e_0$, a good strategy is to compute $h^2_{c,\rm ref}(f)$ for high eccentricity, so that sufficient harmonics are included for accuracy.
We use the same setting of Ref.~\cite{Chen:2016zyo} for a fiducial binary with $\mathcal{M}_0=4.16\times10^8\mathrm{M}_{\odot}$, $z_0 =0.02$, $f^{\rm ref}_{\mathrm{orb},0}=0.1$nHz, $e^{\rm ref}_0=0.9$ but with $j_{\max} = 5\times10^4$.
With these fiducial parameters set, we evaluate~\cref{eqn:hc_delta} and obtain the reference spectrum $h_{c,\mathrm{ref}}(f)$.
We checked the accuracy of this reference spectrum by doubling the maximum number of harmonics to $j_{\max} = 10^5$, which resulted in a relative error of at most $\sim 10^{-5}$. 
In fact, even for a higher initial reference eccentricity of $e_0^{\rm ref} = 0.999$, we obtained a similar relative error when doubling the maximum number of harmonics.

\begin{figure}[!t]
\centering
\includegraphics[width=\columnwidth]{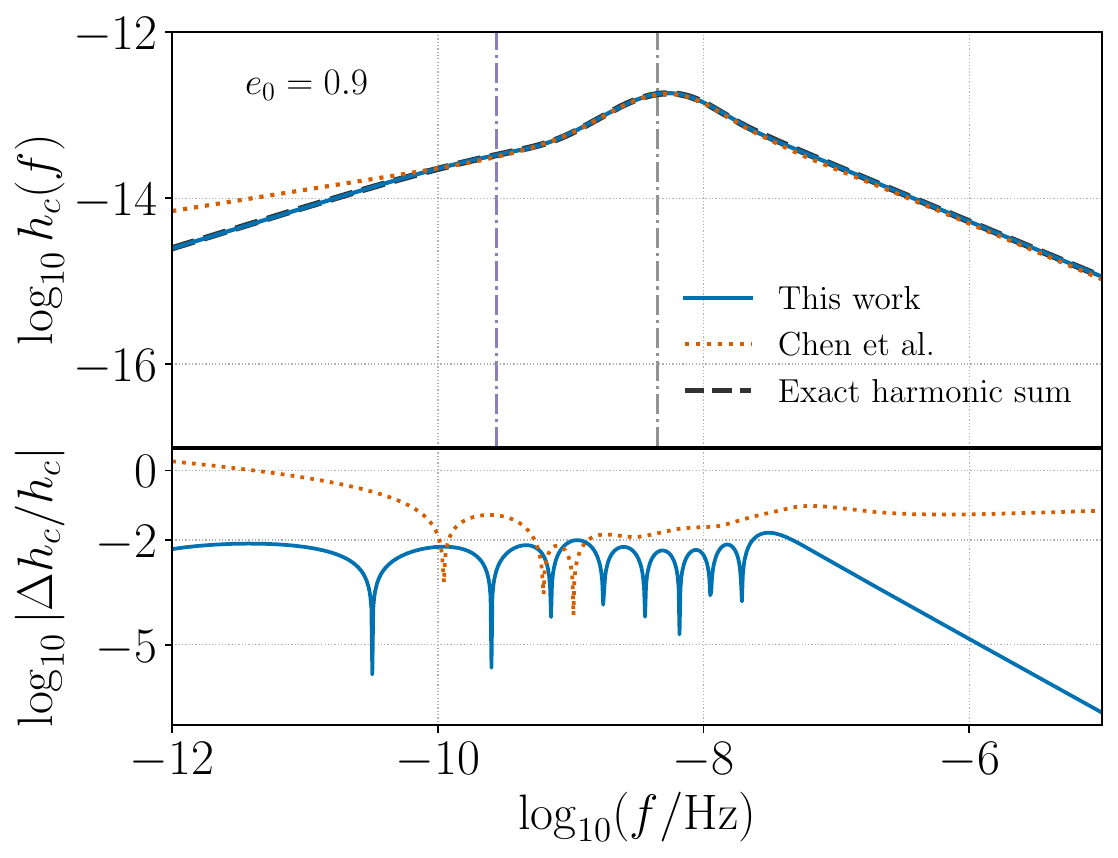}
\caption{
Characteristic strain spectrum $h_{c}(f)$ for the same reference binary described in Ref.~\cite{Chen:2016zyo}.
Dashed lines show the numerical spectra generated as a sum over $j_{\max} = 50000$ harmonics, solid lines correspond to our model predictions, and dotted lines represent the model of Ref.~\cite{Chen:2016zyo}.
Different colors indicate different initial eccentricities $e_{0}$.
The vertical dash-dotted lines indicate the lower (purple) and upper (gray) characteristic frequencies of the bump, $x_l\times10^{-9}\,\mathrm{Hz}$ and
$x_h\times10^{-9}\,\mathrm{Hz}$, respectively.
The lower panel shows the residual $\left| \Delta h_{c} / h_{c} \right|$ as a function of frequency, with the solid and dotted lines corresponding to this work and Ref.~\cite{Chen:2016zyo}, respectively.
}
\label{fig:model_fit}
\end{figure}

In order to fit the reference spectrum $h_{c,\rm ref}^2(f)$, we propose an ansatz that exploits the properties of the function $S(f,f_{\mathrm{orb},0},e_0,z)$.
As pointed out in Ref.~\cite{Huerta:2015pva}, the crucial properties to consider are: 
\begin{align}
    \begin{aligned}
        \lim_{e_0 \rightarrow 0} S(f,f_{\mathrm{orb},0},e_0,z) \rightarrow 1, \\
        \lim _{f \rightarrow \infty} S\left(f,f_{\mathrm{orb},0},e_0,z\right) \rightarrow 1, \\
        \lim _{f \rightarrow 0} S\left(f,f_{\mathrm{orb},0},e_0,z\right)\propto f^{\bar{q}},
    \end{aligned}
\end{align}
where $\bar{q}$ is an exponent that is to be determined.
Further, another important property is that spectrum exhibits a maximum and a dip (see~\cref{fig:model_fit}).
Thus, we propose an ansatz that accounts for these properties:
\begin{subequations}\label{eqn:fit_ansatz}  
\begin{align}
h^2_{c,\mathrm{fit}}(x)&= h^2_{c,\mathrm{circ}}(x) S_{\rm fit} (x), \\
S_{\rm fit} (x) &= \frac{(x/x_{a})^{q_{a}}}{1+(x/x_{a})^{q_{a}}} \times
\left[1+A_{l} \left( \frac{x}{x_l} \right)^{p_{l}}\exp \left( -x/x_l\right)\right] \nonumber \\
& \times \left[1+A_{h} \left (\frac{x}{x_h} \right)^{p_h}\exp \left( -x/x_h\right)\right],
\end{align}  
\end{subequations}
where $x= f/(10^{-9}\mathrm{Hz})$ is the dimensionless frequency.
The structure of the ansatz is a product of a rational power law model that multiplies two exponentially suppressed bump models (terms in the square brackets).
The rational power law dependence captures the low frequency asymptotic power law behavior i.e., $\lim_{x\rightarrow 0} S_{\mathrm{fit}}(x) \propto x^{\bar{q}}$ with an overall amplitude scale set by $x_{a}$.
The terms in the square brackets are characterized by two sets of amplitudes and spectral indices $(A_l,p_l)$ and $(A_h,p_h)$ with corresponding dimensionless frequencies $x_l$ and $x_h$ that determine the location of the bumps in the spectrum.
Importantly, $x_h$ corresponds to the local maximum of the spectrum, while $x_l$ corresponds to the lower frequency dip in the spectrum.
At high frequencies, we have that  $\lim_{x\rightarrow \infty} S_{\mathrm{fit}}(x) = 1$.
Thus, the ansatz correctly captures the power law behavior at low frequencies and asymptotes to unity at high frequency.
In the circular limit $e_0 \rightarrow 0$, one can check that $h_{c}^2(f) \sim h_{c,\rm circ}^2(f)$.
Thus, our ansatz has all the correct properties of the spectrum.

Fitting the ansatz to the reference spectrum, we find the best-fit parameters to be:
\begin{align}
\begin{aligned} \label{eqn:fit_values}
x_{a} &= 5.531, \, & q_{a}  &= 2.942, \\
A_l &= 6.946, \,  & x_l &= 0.2774, \, & p_l &= -0.6217, \\
A_h   &= 4.118, \, & x_h  &= 4.483,  \, & p_h  &= 0.8503.    
\end{aligned}
\end{align}
With these fitting parameters, we find that $\bar{q} = q_a + p_l = 2.3203$. 
In~\cref{fig:model_fit}, we compare our fit to the reference spectrum computed by explicitly summing harmonics up to \( j_{\max} = 5 \times 10^4 \), as well as to the fit proposed in Ref.~\cite{Chen:2016zyo}. 
Across the full frequency range of interest, our fit maintains a maximum relative error of \(1.6\%\). 
In contrast, the fit of Ref.~\cite{Chen:2016zyo} exhibits substantially larger relative errors at low frequencies -- nearly $100 \%$ at $f \sim 10^{-12}$Hz. 
The improvement arises because our ansatz incorporates the correct asymptotic behavior by construction, making it robust across frequency scales. 
By comparison, Ref.~\cite{Chen:2016zyo} obtained their fit by calibrating over a specific frequency interval, and its accuracy therefore deteriorates significantly outside that range.

With the accurate phenomenological fitting model, we can apply it to~\cref{eqn:hc_general} and compute the SGWB by integrating over the differential number density per redshift per chirp mass of the population using
\begin{equation}
\begin{aligned}
h_c^2(f)= & \int d z d \mathcal{M} \frac{d^2 n}{d z d \mathcal{M}}h^{2}_{c, \mathrm{fit}}(f \frac{f_{p,\mathrm{ref}}}{f_{p}}) \\
& \times \left( \frac{f_{p}}{f_{p,\mathrm{ref}}}\right)^{-4/3}\left(\frac{\mathcal{M}}{\mathcal{M}_{0}}\right)^{5/3}
\left(\frac{1+z}{1+z_{0}}\right)^{-1/3}.
\label{eqn:hc_square}
\end{aligned}
\end{equation}

In the following, we translate~\cref{eqn:hc_square} to an integral over the merger rate density per redshift $R_{\rm gw}(z)$ and mass distribution $p(m_1,q)$, which is useful because these can be probed using LVK data.
First, using the chain rule, we have that $dn/dz = (dn/dt_r) (dt_r/dz)$
, where $dz/dt_r = H_0 \mathcal{E}(z) (1+z)$, where $H_0$ is the Hubble constant and $\mathcal{E}(z)$ is the dimensionless Hubble parameter (where we adopt the Planck18 values~\cite{Planck:2018vyg} for the cosmological parameters).
We can then express the comoving number density per proper time $dn/dt_r$ as
\begin{align}
\dfrac{dn}{dt_r} = \int dm_1 dq R_{\rm gw}(z) p(m_1,q),
\end{align}
where $R_{\rm gw}(z)$ is the merger rate density per redshift and $p(m_1,q)$ is the mass distribution.
Using the fact that the quantities $f_p$, $f_{p,\rm ref}$, and $h_{c,\rm fit}^2(f)$ are independent of $m_1, q,$ and $z$, we have 
\begin{equation}
\begin{aligned}\label{eqn:hc_ecc_delta_pop}
&h_c^2(f)=A_{\mathrm{pop}} \times h^{2}_{c, \mathrm{fit}} (f \frac{f_{p,\mathrm{ref}}}{f_{p}}) \left( \frac{f_{p}}{f_{p,\mathrm{ref}}}\right)^{-4/3},
\end{aligned}
\end{equation}
with the dimensionless amplitude $A_{\rm pop}$ given by
\begin{equation}
\begin{aligned}
A_{\mathrm{pop}}  = & \int d z dm_1dq \frac{R_{\mathrm{GW}}(z) p\left(m_1, q\right)}{H_0 \mathcal{E}(z)(1+z)\mathrm{Mpc}^{-3}}\\
& \times \left(\frac{\mathcal{M}}{\mathcal{M}_{0}}\right)^{5/3}
\left(\frac{1+z}{1+z_{0}}\right)^{-1/3}.
\end{aligned}
\end{equation}
Using the same configuration of population described in ~\cite{Chen:2025qyj}, which utilizes the LVK posteriors on the merger rate and mass distribution, we apply Monte Carlo integration to evaluate $A_{\mathrm{pop}}$ and get $A_{\mathrm{pop}}=1.3247 \times10^{-10}$.
For the circular case, the vacuum SGWB spectrum can be parametrized as
\begin{align}
 \Omega_{\mathrm{GW,cir}}(f)=A_{\mathrm{vac}} f^{\gamma}, \label{eqn:hc_circ_pop}  
\end{align}
where $\gamma = 2/3$, and the relation between $A_{\mathrm{pop}}$ and $A_{\mathrm{vac}}$ is given by 
\begin{equation}
\begin{aligned}
A_{\mathrm{vac}} = \frac{\pi^{2/3}}{3\rho_{c}} \frac{\mathcal{M}_{0}^{5/3}\mathrm{Mpc}^{-3}}{(1+z_{0})^{1/3}} A_{\mathrm{pop}}.
\end{aligned}
\end{equation}
In the remainder of this paper, we adopt $\log_{10} A_{\mathrm{vac}} = -10.3036$ (corresponding to $A_{\mathrm{vac}} = 4.97 \times 10^{-11}$), as specified by the ``Rational Power Law'' (RPL) model in~\cite{Chen:2025qyj}.

\begin{figure}[!t]
    \centering
    \includegraphics[width=\columnwidth]{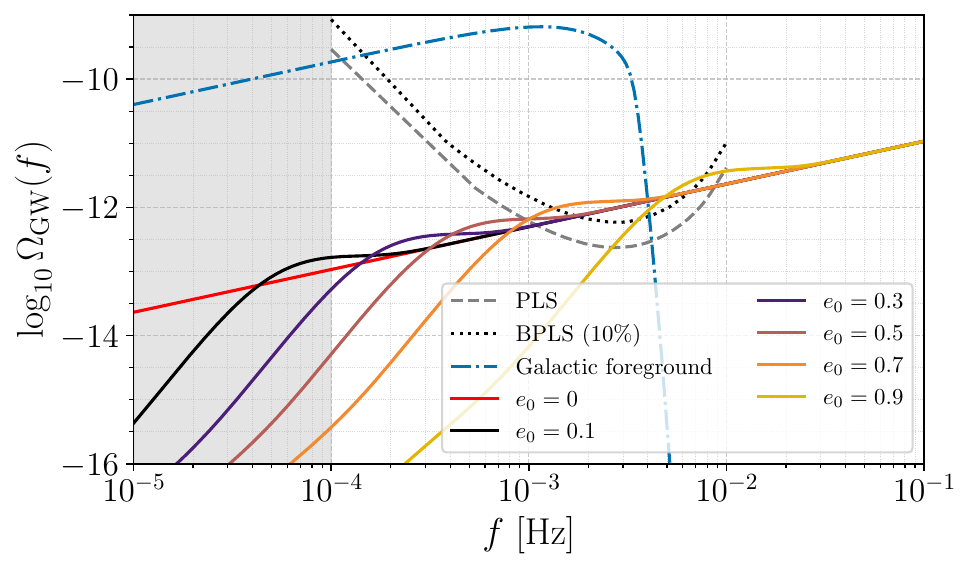}
    \caption{The SGWB spectra $\Omega_{\mathrm{GW}}(f)$ of sBBH for various initial eccentricity $e_{0}$, assuming all binaries have an initial eccentricity $e_{0}$ at $f_{\mathrm{orb},0}=10^{-4}$ Hz. 
    The spectra exhibit a turning point and peak that changes with the eccentricity.
    For comparison, we show the contribution from the galactic white dwarfs confusion noise (blue dash-dotted line).
    The gray dashed curve represents the LISA power-law sensitivity (PLS) for a signal-to-noise ratio of 10.
    The black dotted curve shows the Bayesian power-law sensitivity (BPLS), adopting a Bayes-factor threshold of 10 with a 10\% noise level uncertainty.
    The grey shaded region marks the frequency range outside the LISA sensitivity band.
    }
    \label{fig:ecc_SGWB_dirac}
\end{figure}
In~\cref{fig:ecc_SGWB_dirac}, we show the spectra $\Omega_{\rm GW}(f)$ computed using~\cref{eqn:hc_ecc_delta_pop,eqn:Omega_hc} for $e_0 = \{ 0, 0.1, 0.3, 0.5, 0.7, 0.9 \}$ with $f_{\mathrm{orb},0} = 10^{-4}$Hz.
As expected from previous work such as Refs.~\cite{Huerta:2015pva,Chen:2016zyo}, 
at frequencies below the peak of the spectrum, there is significant power emitted into $j \neq 2$ harmonics, which is why the eccentric spectra are suppressed relative to the circular case (red line), or mathematically $\lim_{f \rightarrow0} \Omega_{\rm GW}(f) \propto f^{2/3+q_{a}}$, with $q_a$ given in~\cref{eqn:fit_values}.
Meanwhile, at frequencies above the peak of the spectrum, most of the power is emitted into the $j=2$ harmonic, which is why it rapidly asymptotes to the circular case, or mathematically $\lim_{f \rightarrow \infty} \Omega_{\rm GW}(f) \sim \Omega_{\rm GW,cir}(f) $.
With increasing eccentricity, observe the peak of the spectrum also shifts to higher frequencies.

The detectability of the signals is strongly determined by the behavior in the mHz regime.
We qualitatively assess this using the BPLS curve (black dotted line) with a Bayes factor threshold of 10, for a reference observation time of $T_{\rm obs} = 4 \mathrm{yr}$ and given a $10\%$ uncertainty on the noise~\cite{Pozzoli:2024hkt}.
For reference, we also show the PLS curve (gray dashed line), which corresponds to a signal-to-noise ratio (SNR) of 10.
We can qualitatively expect to detect stochastic signals when they lie above the BPLS curve in the mHz range.
For the $e_0$ considered, we therefore expect the stochastic signal from eccentric sBBH to be detectable.
In~\cref{fig:ecc_SGWB_dirac}, we also show the stochastic signal produced by unresolved white dwarf binaries in the Milky Way (blue dash-dotted line), which constitutes a Galactic foreground.
Notably, above $e_0 =0.9$, the peak of the spectrum is no longer buried under the Galactic foreground, while it is for lower eccentricities.
In Sec.~\ref{sec:DA}, we perform full Bayesian analysis to assess the detectability of the SGWB from eccentric sBBH and the measurability of the model parameters, while simultaneously also quantifying the impact of the Galactic foreground.

\subsection{Modeling of eccentric SGWB: Thermal distribution for eccentricity}\label{subsec:modeling_ecc_SGWB_thermal}
In general, one needs to consider a distribution of initial eccentricities, in which case, the spectrum is given by
\begin{equation}
\begin{aligned}
h_c^2(f)= & \int d z d \mathcal{M} de_{0} \frac{d^3 n}{d z d \mathcal{M}de_{0}}h^{2}_{c, \mathrm{fit}} (f \frac{f_{p,\mathrm{ref}}}{f_{p}}) \times\\
&\left( \frac{f_{p}}{f_{p,\mathrm{ref}}}\right)^{-4/3}\left(\frac{\mathcal{M}}{\mathcal{M}_{0}}\right)^{5/3}
\left(\frac{1+z}{1+z_{0}}\right)^{-1/3}.
\label{eqn:hc_dndzdMcde0_fit}
\end{aligned}
\end{equation}
Provided that one can factorize the eccentricity dependence as $d^3n/(dz d\mathcal{M} de_0) = d^2 n/ (dz d\mathcal{M}) \times p(e_0)$, where $p(e_0)$ is the probability density for the initial eccentricity, we can translate~\cref{eqn:hc_dndzdMcde0_fit} to an integral over $R_{\rm GW}(z)$ and $p(m_1,q)$:
\begin{align}
\begin{aligned} \label{eqn:hc_ecc_pop}
    h_c^2(f) &= A_{\rm pop} \int de_0 p(e_0) \left( \frac{f_{p} (e_0)}{f_{p,\mathrm{ref}}}\right)^{-4/3} \\
    & \times h^{2}_{c, \mathrm{fit}}\left (f \frac{f_{p,\mathrm{ref}}}{f_{p} (e_0)} \right).    
\end{aligned}
\end{align}
Given an eccentricity distribution $p(e_0)$, one can then readily compute the corresponding SGWB using~\cref{eqn:hc_ecc_pop}.

\begin{figure}
    \centering
    \includegraphics[width=\columnwidth]{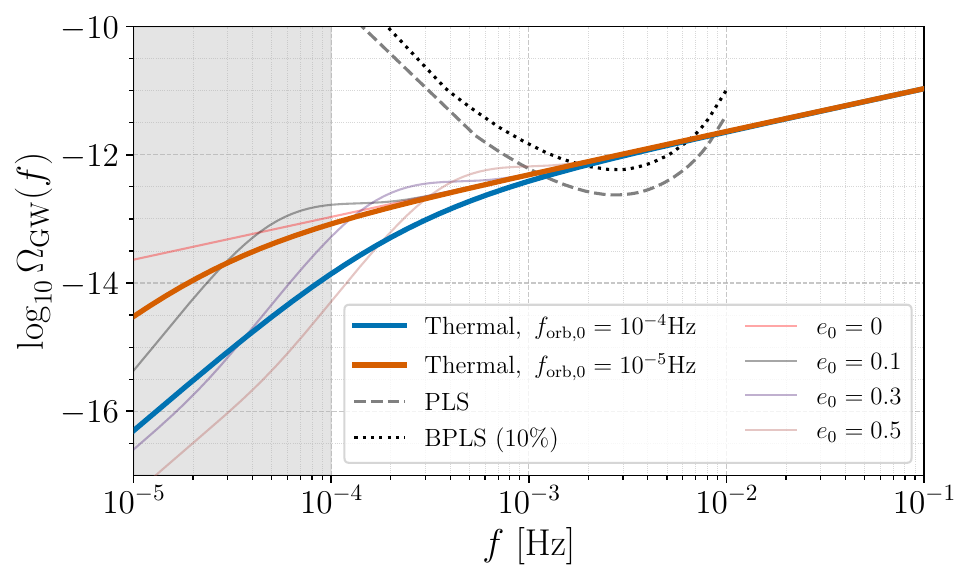}
    \caption{The SGWB spectra $\Omega_{\mathrm{GW}}(f)$ of sBBH with the initial eccentricity $e_{0}$ following a thermal distribution.
    We show the spectra for two values of the initial orbital frequency at which the binaries' eccentricities have thermalized with $f_{\mathrm{orb},0} = \{10^{-5}, 10^{-4} \}$Hz.
    For comparison, we show the spectra corresponding to the Dirac-delta distribution with $e_0 = \{ 0, 0.1, 0.3, 0.5  \}$ with $f_{\mathrm{orb},0} = 10^{-4}$Hz.
    The grey shaded region marks the frequency range outside the LISA sensitivity band.
    }
    \label{fig:ecc_SGWB_thermal}
\end{figure}

In this work, we focus on the ``thermal'' distribution~\cite{Jeans1919,Ambartsumian1937,Heggie1975}, where $p(e_0) = 2 e_0$. 
One can formally derive this distribution with the simple assumption that the phase space number density only depends on the binding energy, and transform the phase space variables to the Delaunay (orbital) elements~\cite{Ambartsumian1937}, after which $p(e_0) = 2 e_0$ falls out as a consequence of the Jacobian.
Notably, triple black hole interactions can also give rise
to a thermal eccentricity distribution, when one of the three black holes is ejected~\cite{Bonetti:2017dan}.
The orbital separation will also follow a corresponding probability distribution, which has been considered in past work on SGWB, such as in Ref.~\cite{Kocsis:2011ch}. 
In our work, we simply assume that all binaries share the same reference orbital frequency $f_{\rm orb,0}$, at which their initial eccentricities are specified.
The value of $f_{\rm orb,0}$, together with the initial eccentricity $e_0$, determines the amount of residual eccentricity retained in the LISA band.
Consequently, increasing $f_{\rm orb,0}$, for a given $e_0$, has a qualitatively similar effect to increasing $e_0$, for a a given $f_{\rm orb,0}$.
We use the thermal eccentricity distribution to study phenomenology of the resulting SGWB, for which the reference orbital frequency $f_{\rm orb,0}$ is the only remaining free parameter.
We leave a more thorough consideration of the astrophysical distributions for the eccentricity and orbital separation for future work.
In ~\cref{fig:ecc_SGWB_thermal}, we show the resulting spectra for two representative values of $f_{\rm{orb},0}$, namely $10^{-5}$ Hz and $10^{-4}$ Hz.
At higher $f_{\rm{orb},0}$, the turnover feature shifts to higher frequencies and the spectrum is more suppressed at lower frequencies.
This is due to the fact that a larger $f_{\rm{orb},0}$ means that binaries have more residual eccentricity when compared at the same orbital frequencies during their evolution.
Thus, with a larger $f_{\mathrm{orb},0}$, the redistribution of power into higher harmonics remains effective up to higher frequencies.
Another consequence of the thermal distribution is that when compared to its Dirac-delta counterparts, the spectrum is shallower at low frequencies i.e., $\lim_{f \rightarrow 0} \Omega_{\rm GW, thermal} (f) \propto f^{2/3+q_{a,\mathrm{ thermal}}}$, with $q_{a,\mathrm{thermal}} < q_{a}$. 
Further, the spectrum transitions more smoothly to the circular model, without exhibiting a strong peak, unlike the Dirac-delta cases.
These features are a result of the fact that the spectrum is integrated over a continuous eccentricity distribution.

\section{Data analysis implications} \label{sec:DA}
We now turn to a comprehensive Bayesian analysis of SGWB signals from eccentric sBBHs in the LISA band.
Specifically, we address the following questions:
\begin{enumerate}[leftmargin=*,itemsep=0pt]
    \item How well can the initial eccentricity be inferred from a LISA detection of the stochastic signal if sBBHs followed a Dirac-delta distribution for $e_0$? 
    What is the impact of the Galactic foreground in inferring such signals?
    Under what conditions is the eccentric model favored over the quasi-circular hypothesis through Bayesian model selection?
    \item If sBBHs eccentricity follows a thermal distribution, are there significant biases when performing parameter estimation with a circular model? Further, can we recover informative posteriors inferring on a model that assumes a Dirac-delta distribution for the eccentricity?
    \item How does including eccentricity in the vacuum SGWB model affect the detection of environmental effects, such as dynamical friction?
\end{enumerate}
In Sec.~\ref{subsec:bayesian}, we provide an overview of the Bayesian methods used. 
We then address each of the above questions in Secs.~\ref{subsec:results_dirac_delta},~\ref{subsec:results_thermal}, and ~\ref{subsec:results_env_ecc} respectively.

\subsection{Bayesian inference of LISA stochastic signals}\label{subsec:bayesian}
We employ the codebase \textsc{Bahamas}~\cite{Bahamas,federicopozzoli_2025_16087705} and adopt the data simulation and inference framework as described in Ref.~\cite{Chen:2016zyo}.
In our analysis, we model the LISA data using time-delay interferometry (TDI) \cite{Tinto:2021}, a technique for suppressing laser noise, and adopt the uncorrelated A, E, and T variables suited for constant, equal-arm configurations. 
Assuming an isotropic, stationary Gaussian background, the stochastic signals in each TDI channel are described in the Fourier domain as a superposition of the power spectral densities from instrumental noise and individual GW source components (i.e., sBBHs and the Galactic foreground).
We coarse-grain the frequency-domain data into segments to reduce computational cost, such that the resulting power spectral density estimators follow a Gamma distribution.
The resulting log-likelihood reads
\begin{align}
\ln \mathcal{L}(\widehat{\boldsymbol{S}} \mid \theta) &=
\sum_{i = 1}^{D} \sum_{k} \left[
  -\ln \Gamma(N)
  - N \ln \left( \frac{S_{k}(f_i; \theta)}{N} \right) \right. \nonumber\\
  &\left. + (N - 1) \ln \widehat{S}_{k}(f_i)
  - \frac{N \, \widehat{S}_{k}(f_i)}{S_{k}(f_i; \theta)}
\right],
\end{align}
where $N$ is the number of neighboring frequency points averaged in each segment, with a total of $D=2024$ segments, and $k=\rm A,\rm E,\rm T$ labels the three TDI channels.
In our work, we restrict the analysis to the $\rm A$ and $\rm E$ channels, only.
The instrumental noise parameters $\theta_n$ describe the spectral amplitudes of the test mass and optical metrology system noises~\cite{QuangNam:2022gjz}.
The Galactic foreground is parametrized by $\theta_{\rm Gal} = \{ \alpha_{\rm Gal}, A_{\rm Gal}, f_{\rm kn}, f_1, f_2\}$, whose values we adopt from Ref~\cite{Karnesisi:2021}.
Recall that the parameters describing the SGWB from eccentric sBBH (with Dirac-delta distribution for the eccentricity) are $\theta_{\rm sBBH} = \{ A_{\rm vac}, \gamma, e_0 \}$.
Thus, when performing parameter estimation with our eccentric model, the parameters we sample on are $\theta = \theta_{\rm sBBH} \cup \theta_{\rm Gal} \cup \theta_n$.
For $\theta_n$ and $\theta_{\rm Gal}$, we adopt the same priors as used in our previous work~\cite{Chen:2025qyj}.

\subsection{Implications of SGWB produced with Dirac-delta eccentricity distribution}{\label{subsec:results_dirac_delta}}

For a quantitative evaluation of measurability, we consider a Dirac-delta eccentricity distribution in which all binaries share the same initial eccentricity $e_0$, and perform injections using the eccentric SGWB model given by~\cref{eqn:hc_ecc_delta_pop}. 
We consider injections with initial eccentricities $e_0 \in [0.1$,$0.9]$ (with linear spacing of 0.1), while fixing $f_{\mathrm{orb},0}=10^{-4}$ Hz to isolate the effect of the initial eccentricity.
These injections correspond to ``residual'' eccentricities ranging from $e_{f_{\mathrm{orb},0} = 10 \mathrm{Hz}} \sim 5.3 \times 10^{-7}$ to $\sim 7.7 \times 10^{-5}$, when evaluated at an orbital frequency of $f_{\mathrm{orb}}=10$ Hz, indicating that the binaries are effectively circularized in the LVK band.
We perform parameter estimation on these injected signals using the same eccentric SGWB model as described by~\cref{eqn:hc_ecc_delta_pop}.
In order to obtain informative posteriors on $e_0$, we fix the spectral index $\gamma$ that enters~\cref{eqn:hc_ecc_delta_pop} to $\gamma = 2/3$, i.e. the value for the circular SGWB model.
We also separately recover with the circular (power law) SGWB model, described by~\cref{eqn:hc_circ_pop}, but here we sample on $\gamma$ to perform model selection, and assess at what eccentricity the eccentric model is strongly preferred.

We quantify the preference of the eccentric model over the circular (power law) model, by computing the Bayes factor $\log_{10}\mathcal{B}^{\rm ecc}_{\rm cir} = \log_{10}(\mathcal{Z}_{\rm ecc}/\mathcal{Z}_{\rm cir})$, where $\mathcal{Z}_{\rm ecc}$ and $\mathcal{Z}_{\rm cir}$ denote the Bayesian evidences for the respective models. 
A value of $\log_{10} \mathcal{B}_{\rm cir}^{\rm ecc} > 1$ is interpreted as strong evidence in favor of the eccentric model. 
For all analyses reported below, we use the nested sampling routine available in \textsc{Bahamas}, as implemented in \textsc{Nessai}~\cite{nessai,Williams:2021qyt,Williams:2023ppp} . 
Specifically, we used \texttt{nlive} $=2000$, \texttt{dlogz} $= 0.1$.
We tested the convergence and robustness of our sampler by increasing \texttt{nlive}, and found that \texttt{nlive} $=2000$ suffices for our work.
Given that we are only interested in evidence estimates that result in log Bayes factors $ \gtrsim 1$, we found it sufficient to set \texttt{dlogz} $=0.1$.

\begin{table}[!b]
\centering
\begin{ruledtabular}
\begin{tabular}{c|c|c|c|c|c}
$e_{0}^{\rm inj}$ 
& $e_{f_{\rm orb} = 10 \,\mathrm{Hz}}$
& SNR
& $\delta e_{0}/e_{0}^{\rm inj}$ 
& $\delta A_{\rm vac}/A_{\rm vac}^{\rm inj}$ 
& $\log_{10}\mathcal{B}^{\mathrm{ecc}}_{\mathrm{cir}}$ \\
\hline
$0.1$ & $5.372\times10^{-7}$ & $200$ & $3.069$ & $0.05113$ & $0.7360$ \\
$0.2$ & $1.136\times10^{-6}$ & $200$ & $1.557$    & $0.05049$ & $0.9060$ \\
$0.3$ & $1.877\times10^{-6}$ & $200$ & $0.9955$    & $0.05064$ & $0.6035$ \\
$0.4$ & $2.886\times10^{-6}$ & $201$ & $0.7568$   & $0.05172$ & $0.7175$ \\
$0.5$ & $4.402\times10^{-6}$ & $201$ & $0.6461$   & $0.05049$ & $0.8518$ \\
$0.6$ & $6.952\times10^{-6}$ & $206$& $0.5232$   & $0.05081$ & $0.9513$ \\
$0.7$ & $1.193\times10^{-5}$ & $228$ & $0.4814$   & $0.05650$ & $0.8157$ \\
$0.8$ & $2.437\times10^{-5}$ & $282$ & $0.03269$  & $0.1073$  & $0.2662$ \\
$0.9$ & $7.783\times10^{-5}$ & $181$ & $0.003499$ & $0.06279$ & $11.12$ \\
\end{tabular}
\end{ruledtabular}
\caption{Posterior statistical uncertainties (from $90\%$ quantiles) on $e_{0}$ and $A_{\rm vac}$, together with the Bayes factor between the eccentric and circular hypotheses. 
The second column lists the eccentricity evolved to $f_{\rm orb} = 10\,\mathrm{Hz}$. 
Results are marginalized over Galactic foreground parameter uncertainties.
}
\label{tab:posterior_statistical_uncertainties_delta_e0}
\end{table}

\begin{figure*}[t]
    \centering
    \includegraphics[width=0.7\linewidth]{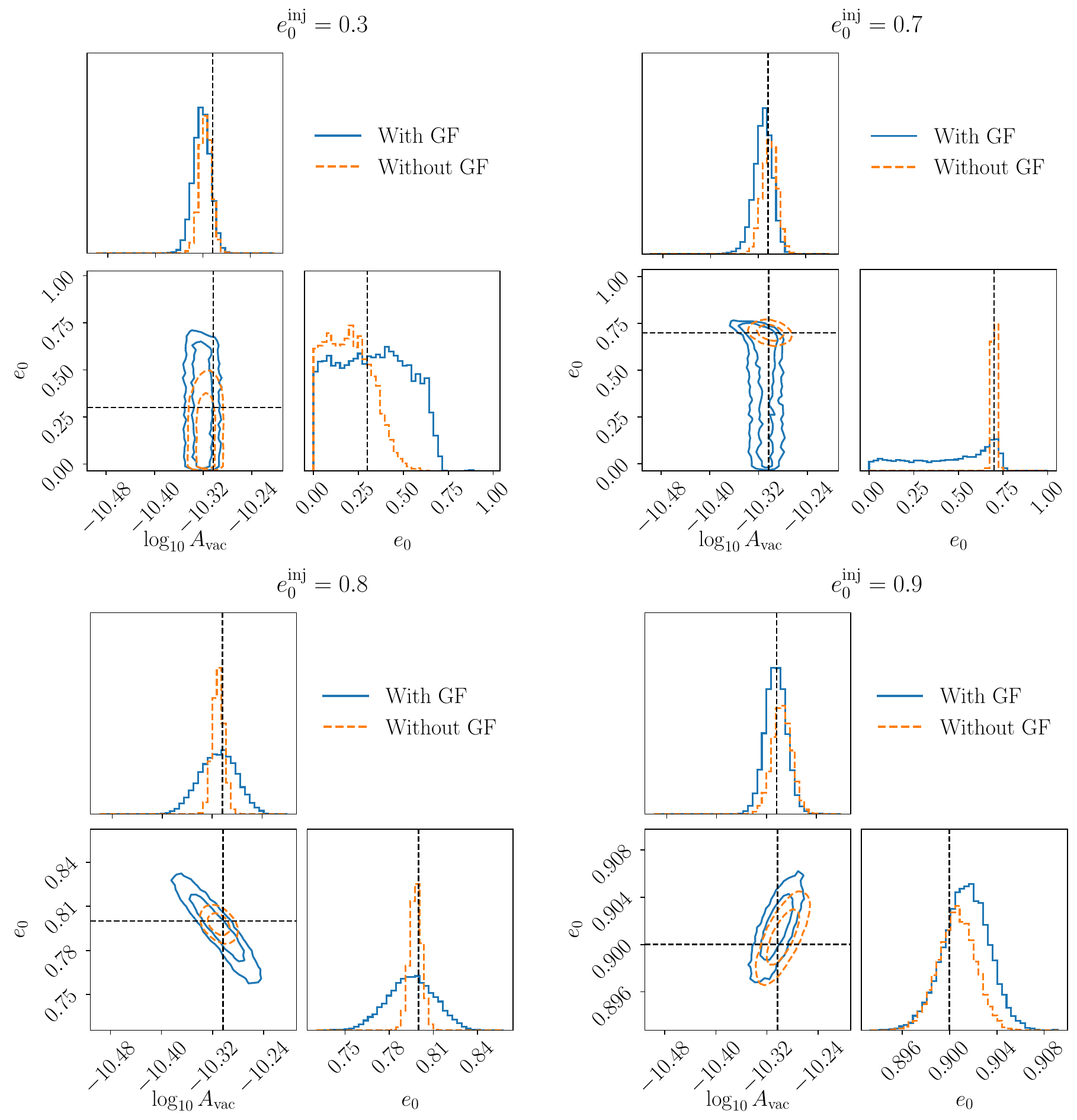}
    \caption{Marginalized posteriors for the eccentric model across various initial eccentricity $e_{0}$. 
    Solid blue (dashed orange) contours and histograms correspond to analyses marginalized over the(assuming the true) Galactic foreground parameters.
    }
    \label{fig:ecc_template_recovery}
\end{figure*}

\begin{figure}[t]
    \centering
    \includegraphics[width=\columnwidth]{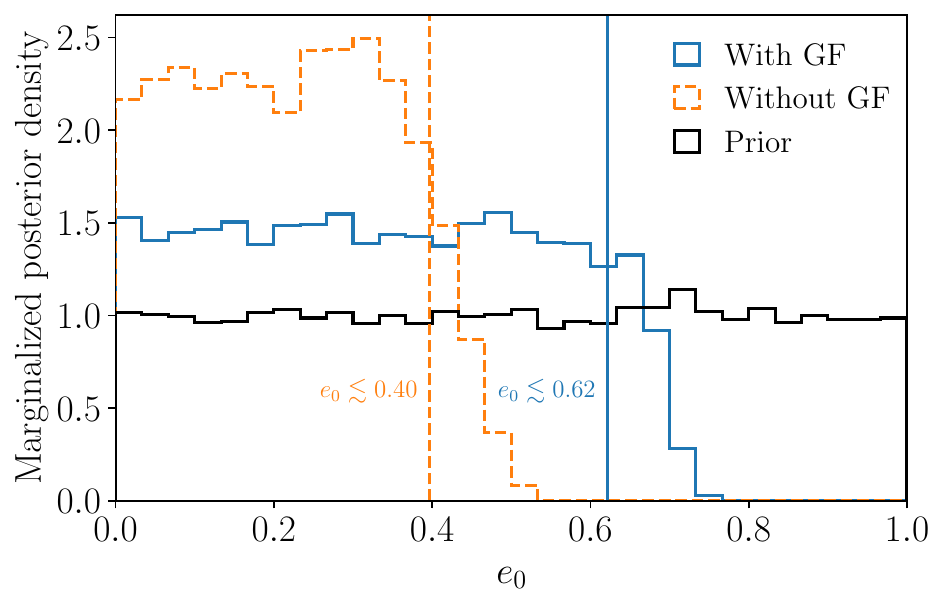}
    \caption{
    Posteriors on $e_0$ (with $f_{\mathrm{orb},0} = 10^{-4}$Hz) obtained with (solid, blue) and without (dashed, orange) including the galactic foreground parameters, given a LISA stochastic signal produced by quasi-circular sBBHs.
    The $90\%$ right-sided credible intervals are shown with vertical lines.
    For reference, we show the adopted uniform prior (solid, black) on $e_0$.
    }
    \label{fig:ecc_upper_bound}
\end{figure}
For each injected signal, we compute statistical uncertainties from the 90\% credible interval widths, defined as $\delta \theta/\theta^{\rm inj}=\left|\theta^{95 \%}-\theta^{5 \%}\right|/(2\theta^{\rm inj})$ for $\theta \in \{A_{\mathrm{vac}}, e_0\}$, along with the injected SNR, and the log Bayes factor $\log_{10} \mathcal{B}^{\rm ecc}_{\rm cir}$, which we show in~\cref{tab:posterior_statistical_uncertainties_delta_e0}. 
For a set of representative injected eccentricities, \cref{fig:ecc_template_recovery} shows the marginalized posterior distributions of $A_{\rm vac}$ and $e_0$, comparing results obtained with (solid lines) and without (dashed lines) uncertainties in the Galactic foreground parameters. In the latter case, the Galactic parameters $\theta_{\rm Gal}$ are fixed to their injected values, effectively corresponding to a Dirac delta prior.

In all cases, we obtain informative posteriors on $A_{\rm vac}$, with $\delta A_{\rm vac} / A_{\rm vac}^{\rm inj} \lesssim 0.1$.
Instead, we only obtain informative posteriors on $e_0$ for injected values $e^{\rm inj}_0 \geq 0.8$.
For example, with $e^{\rm inj}_0 = 0.3$, as shown in~\cref{fig:ecc_template_recovery}, the posteriors remain uninformative even without including the Galactic foreground parameters, with strong support at $e_0 = 0$. 
At $e^{\rm inj}_0 = 0.7$, the posterior exhibits a clear peak near the injected value, with a smaller statistical uncertainty when excluding the Galactic foreground parameters.
As the injected eccentricity increases, the impact of the Galactic foreground also diminishes, due to its suppressed contribution above $\sim 3 \times 10^{-3}$Hz, as shown earlier in~\cref{fig:ecc_SGWB_dirac}.
For the case of $e_0 = 0.9$, the sBBH spectrum has a peak that is well above the frequency where the Galactic foreground is suppressed, and thus the posteriors on $e_0$ are nearly identical whether or not the Galactic foreground parameters are included. 
We find that in all cases, the eccentric model is marginally preferred over the circular model, with the log Bayes factors in the range of $\log_{10} \mathcal{B}^{\rm ecc}_{\rm cir} \sim 0.26$---$0.90$ when $e_0 \in [0.1,0.8]$.
Meanwhile, when $e_0 = 0.9$, there is very strong evidence against the circular model, with $\log_{10} \mathcal{B}^{\rm ecc}_{\rm cir} \approx 11$.
In Appendix~\ref{app:additional}, we provide additional analysis in terms of the posterior predictives for both the eccentric and circular models, and further explain the trends observed in~\cref{tab:posterior_statistical_uncertainties_delta_e0}.

In~\cref{fig:ecc_upper_bound}, we show the constraints on $e_0$ obtained with our model, given a detection of a stochastic signal from quasi-circular sBBHs.
Specifically, we inject a circular SGWB described 
by~\cref{eqn:hc_circ_pop}, and obtain marginalized posteriors on $e_0$ with our eccentric model described by~\cref{eqn:hc_ecc_delta_pop}, with (blue solid histogram) and without (orange dashed histogram) including the Galactic foreground parameters.
We find a constraint (90 $\%$ upper credible interval) of $e_0 \lesssim 0.62$ when including the Galactic foreground parameters, which translates to a residual eccentricity of $e_{f_{\mathrm{ orb}} = 10 \mathrm{Hz}} \sim (\mathrm{a \ few}) \times 10^{-6}$.
This constraint is consistent with results shown in~\cref{fig:ecc_template_recovery,tab:posterior_statistical_uncertainties_delta_e0}.

\begin{figure*}[!t]
    \centering
    \includegraphics[width=1.0\linewidth]{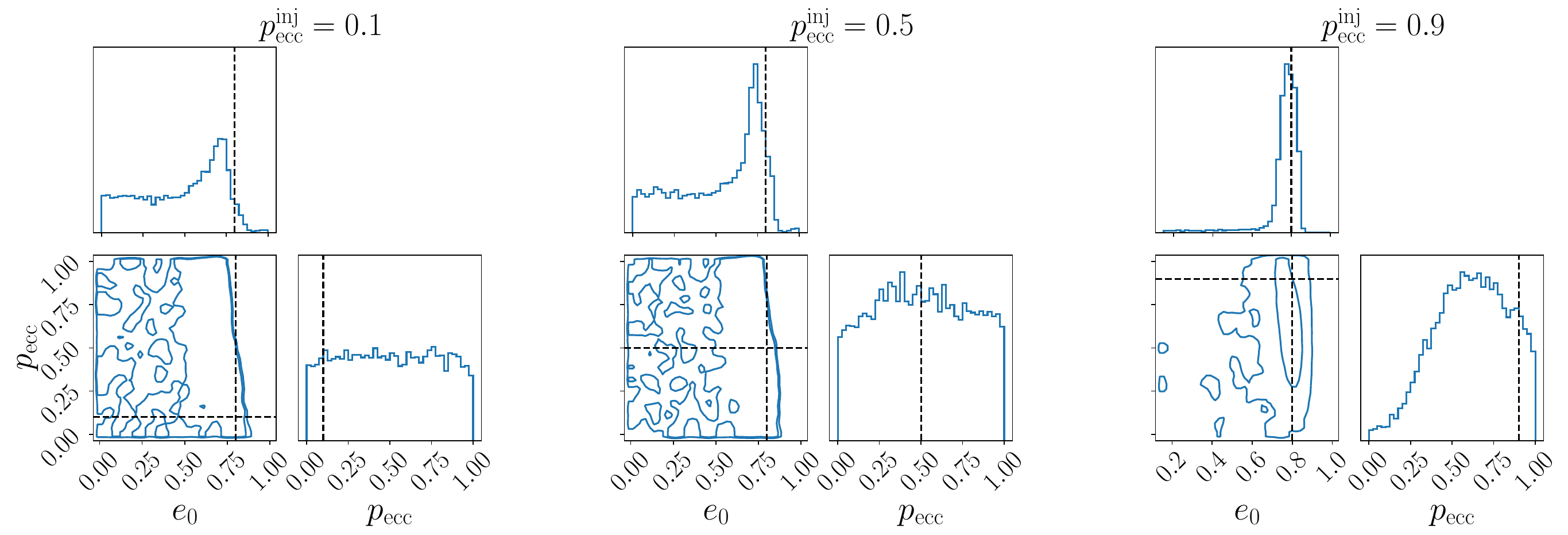}
    \caption{
    Marginalized posteriors on $e_0$ and $p_{\rm ecc}$ inferred with the mixture model.
    In all cases, the injected spectra are generated with $e^{\rm inj}_0 = 0.8$ at $f_{\rm{ orb},0} = 10^{-4}$Hz, while the injected values of $p_{\rm ecc}$ are $0.1$ (left), $0.5$ (middle), and $0.9$ (right), and shown as dashed black lines in each panel.
    }
    \label{fig:fraction_injection_recovery}
\end{figure*}

We now relax the assumption of binaries with identical eccentricity and study how the results change when only a fraction of the spectrum originates from eccentric binaries.
Prior to analyzing the case of the thermal eccentricity distribution (discussed below in Sec.~\ref{subsec:modeling_ecc_SGWB_thermal}), we now partially address this by adopting a phenomenological mixture model (similar to our previous work in Ref.~\cite{Chen:2025qyj}) parametrized by a fraction $p_{\mathrm{ecc}}$,
\begin{equation}
\label{eqn:omega_frac}
\Omega_{\mathrm{Frac}} = p_{\mathrm{ecc}} \Omega_{\mathrm{ecc}}+(1-p_{\mathrm{ecc}})\Omega_{\mathrm{cir}},
\end{equation}
where $\Omega_{\mathrm{ecc}}$ is generated assuming a Dirac-delta distribution for the eccentricity, and $\Omega_{\mathrm{cir}}$ corresponds to the vacuum circular SGWB spectrum. 
We injected signals using~\cref{eqn:omega_frac} with $e^{\rm inj}_0 = 0.8$ for $p_{\rm ecc}^{\rm inj} \in \{0.1,0.3,0.5,0.7,0.9\}$, and performed parameter estimation with the mixture model by also sampling on $p_{\rm ecc}$, and separately with the circular model (which corresponds to either fixing $e_0 = 0$ or $p_{\rm ecc} = 0$ in the mixture model).
In~\cref{fig:fraction_injection_recovery}, we show the one and two dimensional marginalized posteriors on $e_0$ and $p_{\rm ecc}$ for a few representative cases.
In~\cref{tab:fraction_bayes_factors}, we show the log Bayes factor between the mixture and circular models for each case.
Going forward, we only report analyses with the Galactic foreground parameters inferred upon, being a more realistic scenario.

For small fractions such as $p_{\rm ecc} = 0.1$ or $p_{\rm ecc} = 0.3$, there is only a mild peak for the marginalized posterior $e_0$, which is close to the injected value of $e_0 = 0.8$. 
Meanwhile, the marginalized posteriors on $p_{\rm ecc}$ are entirely uninformative for these cases.
However, for a larger injected fraction of $p_{\rm ecc} = 0.9$, as shown in~\cref{fig:fraction_injection_recovery}, the marginalized posteriors on $e_0$ and $p_{\rm ecc}$ are informative, with the latter nearly consistent with what we had obtained before in~\cref{fig:ecc_template_recovery}.
This is consistent with what we find for the Bayes factors -- the evidence in favor of the mixture model increases with $p_{\rm ecc}$, with the circular model strongly disfavored for $p_{\rm ecc} = 0.9$.
In other words, even with the Dirac-delta distribution with a large eccentricity of $e_0 = 0.8$, one further needs a large mixing fraction to identify the effects of eccentricity.

\begin{table}[!t]
\centering
\caption{Log Bayes factors $\log_{10}\mathcal{B}_{\mathrm{cir}}^{\mathrm{frac}}$ comparing the mixture model to the circular model, for different injected values of $p_{\rm ecc}$.
}
\label{tab:fraction_bayes_factors}
\begin{ruledtabular}
\begin{tabular}{lccccc}
$p_{\rm ecc}$ & 0.1   & 0.3   & 0.5   & 0.7   & 0.9  \\
\hline
{$\log_{10}\mathcal{B}_{\mathrm{cir}}^{\mathrm{frac}}$}    & -0.4034 & -0.2954 & 0.4516 & 0.9786 & 4.735 \\
\end{tabular}
\end{ruledtabular}
\end{table}

\subsection{Implications of SGWB produced with thermal eccentricity distribution}{\label{subsec:results_thermal}}
We now analyze the spectrum produced by sBBH whose eccentricities follow the thermal distribution, as discussed earlier in Sec.~\ref{subsec:modeling_ecc_SGWB_thermal}. 
We specifically consider two cases with $f_{\rm orb} = 10^{-5}$Hz and $f_{\rm orb} = 10^{-4}$Hz, where the latter case results in a spectrum that is more suppressed due to larger residual eccentricities in the LISA band.

\begin{figure}
    \centering
    \includegraphics[width=\linewidth]{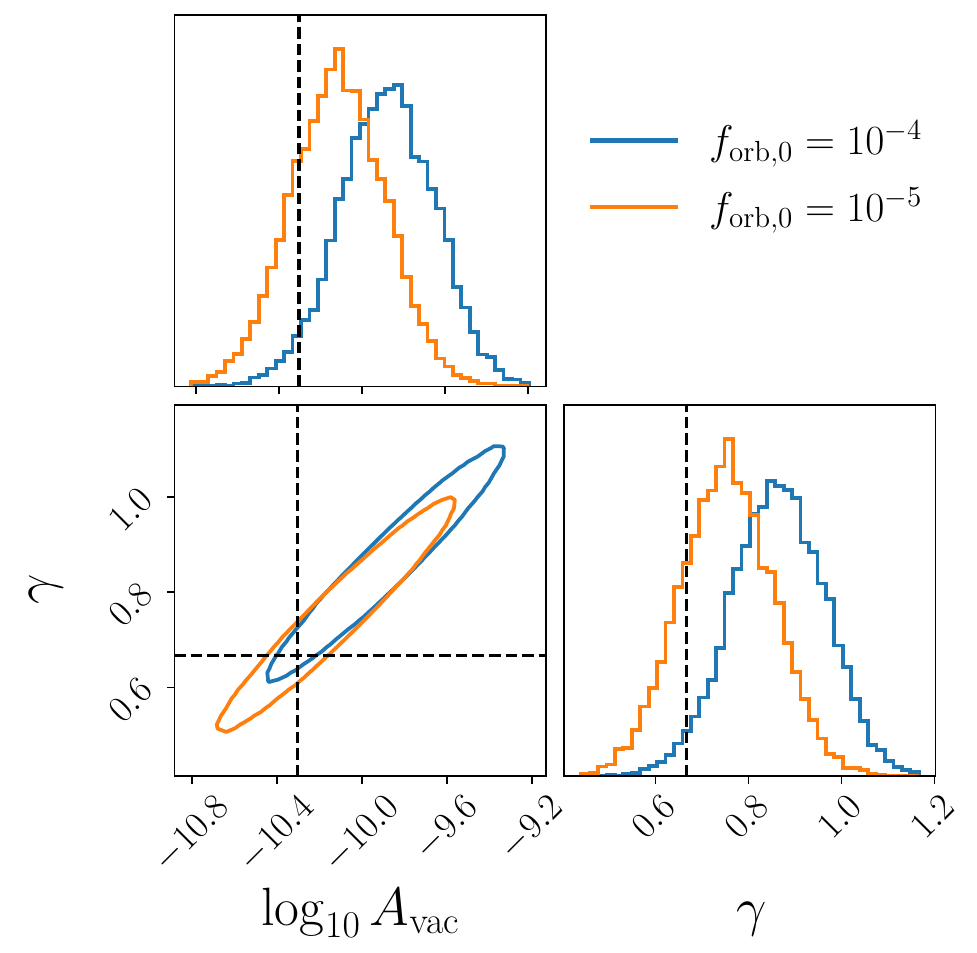}
    \caption{Biases in the circular model when analyzing a stochastic signal from sBBHs with eccentricity following a thermal distribution.
    }
    \label{fig:thermal_bias}
\end{figure}

First, we address the significance of the systematic biases when using the circular model, given each injected spectrum.
In~\cref{fig:thermal_bias}, we show the marginalized posteriors (blue for $f_{\rm orb} = 10^{-4}$Hz and orange for $f_{\rm orb} = 10^{-5}$Hz) on the circular model parameters $A_{\rm vac}$ and $\gamma$, with the true values of the circular model shown by the dashed lines.
For the case with $f_{\rm orb} = 10^{-4}$Hz, we find that the systematic errors in the one dimensional marginalized posteriors are comparable to the respective statistical errors, while they are smaller for the case with  $f_{\rm orb} = 10^{-5}$Hz.
Thus, if the sBBHs formed at lower frequencies of $f_{\rm orb} \lesssim 10^{-5}$Hz, with their eccentricities following a thermal distribution, the resulting SGWB would effectively be consistent with a circular model in the LISA band.
Further, note that the spectral index is biased to larger values than the circular case, while the opposite holds true for the vacuum.
This is because the spectrum is suppressed at lower frequencies with a larger spectral index.

\begin{figure}
    \centering
    \includegraphics[width=\linewidth]{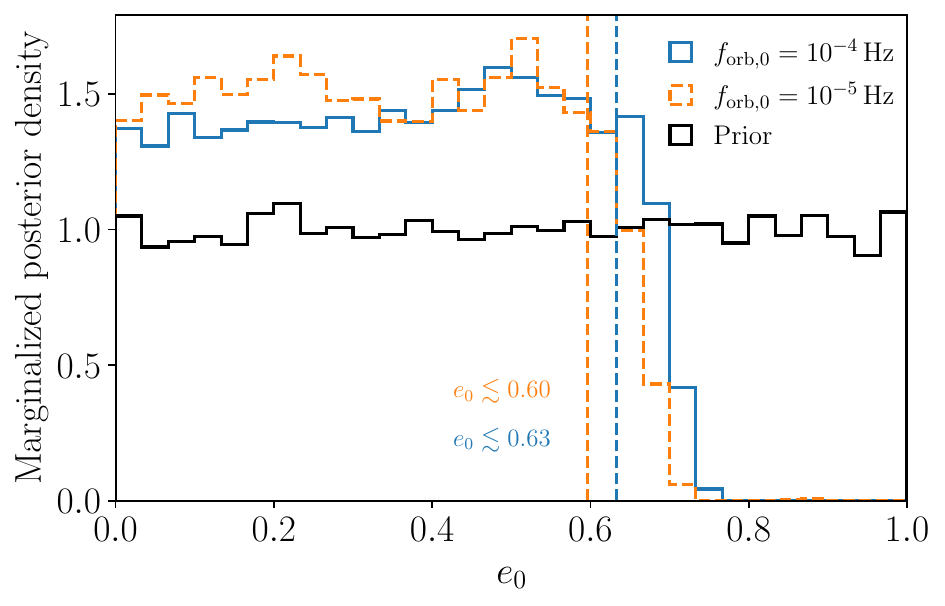}
    \caption{Constraint on $e_0$ when recovering with the parametric model, given a SGWB that is generated by binaries following a thermal distribution of initial eccentricity.}
    \label{fig:thermal_constraint}
\end{figure}

If we instead use the eccentric model described by the Dirac-delta distribution [cf.~\cref{eqn:hc_ecc_delta_pop}], can we obtain any informative posterior for a nonzero $e_0$ from the thermal spectra? 
In such a case, the inferred $e_0$ would represent an effective eccentricity, and we can assess whether the
two cases of the injected spectra (with $f_{\rm orb} = 10^{-5}$Hz and $f_{\rm orb} = 10^{-4}$Hz respectively) lead to different posterior constraints on this effective eccentricity.
However, as shown in~\cref{fig:thermal_constraint}, the inferred posteriors on $e_0$ have strong support at $e_0 = 0$, with constraints (from the upper $90\%$ credible interval) of $e_0 \lesssim 0.6$ in both cases.
As expected the eccentric model with the Dirac-delta distribution is only informative with a nonzero peak for $e_0$ when the injected $e_0 \gtrsim 0.7$ due to the peak / turning point of the spectrum is $\sim 10^{-3}$Hz, where LISA is most sensitive.
The spectrum resulting from the thermal distribution, in both cases of $f_{\rm orb} = 10^{-5}$Hz and $f_{\rm orb} = 10^{-4}$Hz, has a turning point that is below this sensitive region, and is therefore similar to injecting with $e_0 < 0.7$, resulting in posteriors largely supporting $e_0 = 0$.

\subsection{Distinguishing environmental effects from an eccentric vacuum model}
{\label{subsec:results_env_ecc}}
\begin{figure*}[!t]
    \centering
    \includegraphics[width=0.95\linewidth]{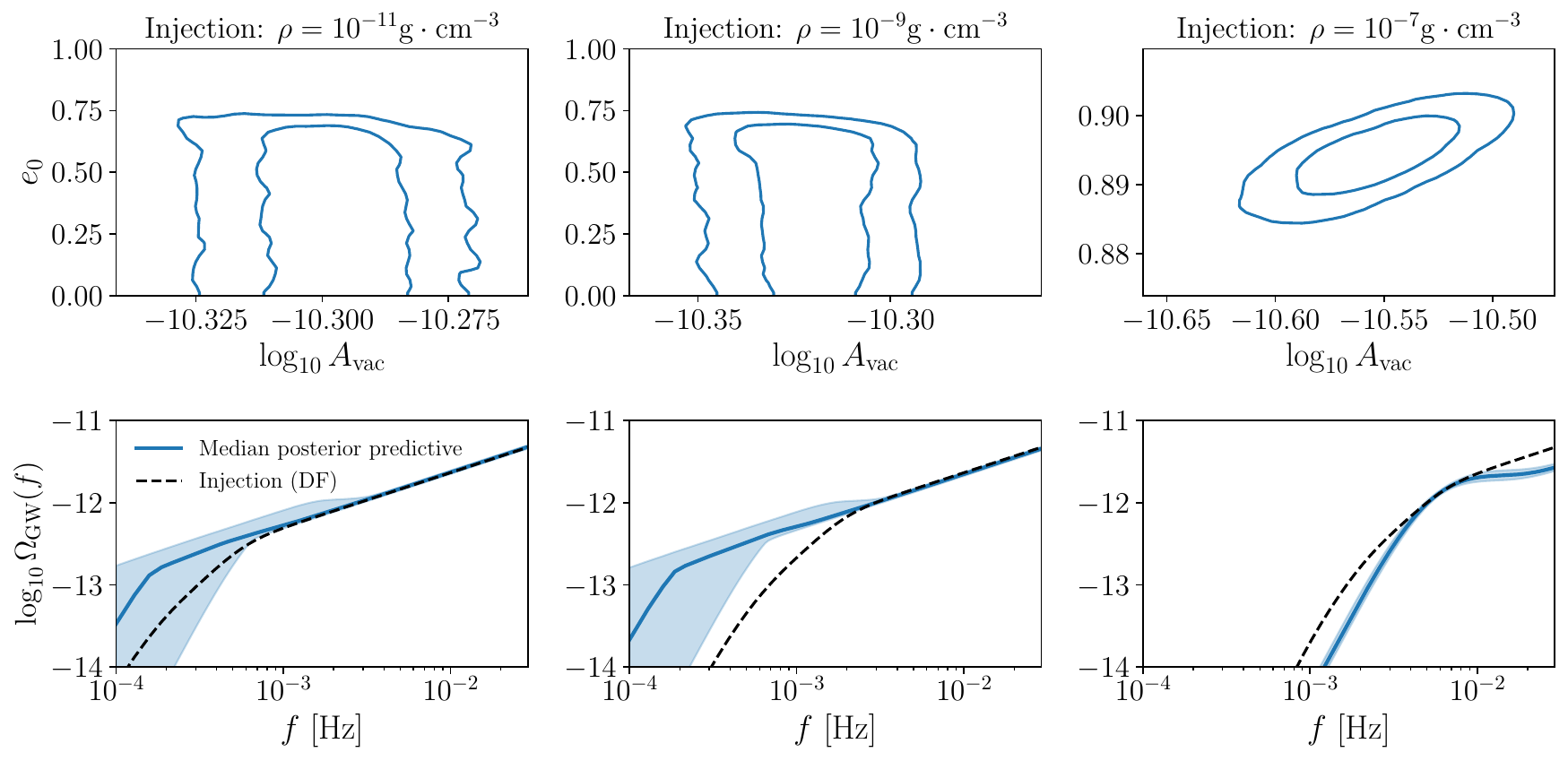}
    \caption{
    Recovery of injected stochastic signals undergoing dynamical friction using the eccentric model for three representative gas densities, $\rho=10^{-11}, 10^{-9}, 10^{-7}\mathrm{g\,cm^{-3}}$.
    The top panels show the two marginalized posteriors on ($\log_{10} A_{\rm vac}, e_0$)
    , and the bottom panels show the corresponding posterior predictive spectra, where the solid blue curves denote the posterior medians, the shaded regions indicate the $90\%$ credible intervals, and the black dashed curves represent the injected signals.
    }
    \label{fig:df_injet_ecc_recovery}
\end{figure*}

In our previous work~\cite{Chen:2025qyj}, we studied the impact of astrophysical environments on the SGWB from quasi-circular sBBHs.
Since astrophysical environments typically introduce additional energy dissipation into the
binary evolution, they lead to characteristic distortions of the SGWB spectrum.
Based on asymptotic behavior of these spectral modifications, we developed a generic phenomenological parametrization, referred to as the ``Rational Power Law + Peak'' (RPLP) model.
The RPLP model is sufficiently flexible to describe a broad class of environmental effects.
We found that the effects of gas dynamical friction acting on binaries embedded in a gaseous medium can be measured, and that the vacuum model can be rejected for typical disk gas densities.
We now address how those results are affected by allowing for nonzero $e_0$ in the vacuum model.
Specifically, we inject spectra from quasi-circular sBBH undergoing dynamical friction for disk gas densities $\rho^{\rm inj} \in \{ 10^{-11}, 10^{-10} , 10^{-9}, 10^{-8} , 10^{-7} \} \mathrm{g\,cm^{-3}}$.
For each injected spectrum, we perform two sets of parameter estimations, one with the dynamical friction model (same as injection) and another with a vacuum SGWB model, but allowing for the effects of eccentricity that follows a Dirac-delta distribution, as described by~\cref{eqn:hc_ecc_delta_pop}.

In the top panel of~\cref{fig:df_injet_ecc_recovery}, for a few representative cases, we show the two dimensional marginalized posteriors on $A_{\rm vac}$ and $e_0$, while in the bottom panel, we show the posterior predictive corresponding to the eccentric model.
To compute the posterior predictive, we evaluate the eccentric model power spectral density (PSD) from each posterior samples and at each frequency, and show the corresponding median and $90\%$ credible intervals.
Meanwhile, in~\cref{tab:DF_ecc_bayes_factors}, we show the Bayes factors between the dynamical friction model and the vacuum models, with the top row corresponding to the eccentric vacuum case, and the bottom corresponding to the quasi-circular vacuum case (which we reproduced from our previous work~\cite{Chen:2025qyj}).

\begin{table}[!b]
\centering
\caption{Log Bayes factors comparing the dynamical friction model (RPLP) to the vacuum models, for different values of $\rho^{\rm inj}$. The top row is for eccentric vacuum case, and the bottom row is for the quasi-circular vacuum case.}
\label{tab:DF_ecc_bayes_factors}
\begin{ruledtabular}
\begin{tabular}{lccccc}
$\rho\ (\mathrm{g\,cm^{-3}})$ & $10^{-11}$ & $10^{-10}$ & $10^{-9}$ & $10^{-8}$ & $10^{-7}$ \\
\hline
{$\log_{10}\mathcal{B}_{\mathrm{ecc}}^{\mathrm{DF}}$}  & -0.01275 & 0.01033 & -0.09733 & -0.1328 & 1.337\\
{$\log_{10}\mathcal{B}_{\mathrm{cir}}^{\mathrm{DF}}$}  & 1.397 & 1.375 & 1.407 & 1.012 & 1.474\\
\end{tabular}
\end{ruledtabular}
\end{table}

When including eccentricity, the evidence in favor of the dynamical friction model decreases (compared to the circular model) in all cases.
Only for a large disk gas density of $\rho^{\rm inj} = 10^{-7} \mathrm{g\,cm^{-3}}$, we can strongly reject the eccentric vacuum hypothesis.
For lower values of the disk gas densities, the log Bayes factors are inconclusive, given that $\log \mathcal{B}^{\rm DF}_{\rm ecc} \approx 0$ to within the uncertainties of estimating the evidences.
This is also seen from the marginalized posteriors and posterior predictives of the eccentric model shown in~\cref{fig:df_injet_ecc_recovery}.
For smaller disk gas densities, such as $\rho^{\rm inj} = \{ 10^{-11}, 10^{-9} \} \mathrm{g\,cm^{-3}}$, the inference of $e_0$ is uninformative, with large posterior predictive uncertainties at low frequencies.
With $\rho^{\rm inj} = 10^{-7} \mathrm{g\,cm^{-3}}$,
there is a clear nonzero peak for $e_0$, and the corresponding posterior predictive is distinguishable from the injected spectrum.
Moreover, for this case, we have $\log_{10} \mathcal{B}^{\rm DF}_{\rm ecc} \approx 1.3$, which exceeds our adopted threshold for strong Bayesian evidence ($\log_{10}\mathcal{B}>1$).
For this density, our Bayes factor is consistent with what we had found when assuming no eccentricity for the vacuum model.
Thus, for such large values of the disk gas density, we can strongly reject the vacuum model even when allowing for eccentricity.

\section{Constraining the maximum eccentricity in the band of ground-based detectors with LISA}\label{sec:erefmax}
So far, our analyses have focused on eccentricity distributions defined at a LISA-band reference frequency, but it is natural to ask how LISA observations of the SGWB can constrain the eccentricity of the sBBH population when the latter is characterized by $e_{20\,{\rm Hz}}$, namely the eccentricity defined at a ground-based detector reference GW frequency of 20 Hz for the dominant $(2,2)$ mode~\cite{Shaikh:2023ypz}.
To this end, we consider an illustrative case where the population of sBBHs whose eccentricity at $f_{\rm orb}=10\,\mathrm{Hz}$ follows a uniform distribution between $0$ and a maximum value $e^{\mathrm{max}}_{20\,\mathrm{Hz}}$, which is the single population hyperparameter we aim to constrain.
Such a distribution is motivated by the typical priors used in LVK and related analyses~\cite{Romero-Shaw:2021ual,Bonino:2022hkj,Gupte:2024jfe}.
For each value of $e^{\mathrm{max}}_{20\,\mathrm{Hz}}$, we evolve the population backward to the LISA band and compute the corresponding SGWB spectrum using the eccentric model developed in~\cref{subsec:modeling_ecc_SGWB}.

\begin{figure}[!t]
    \centering
    \includegraphics[width=\columnwidth]{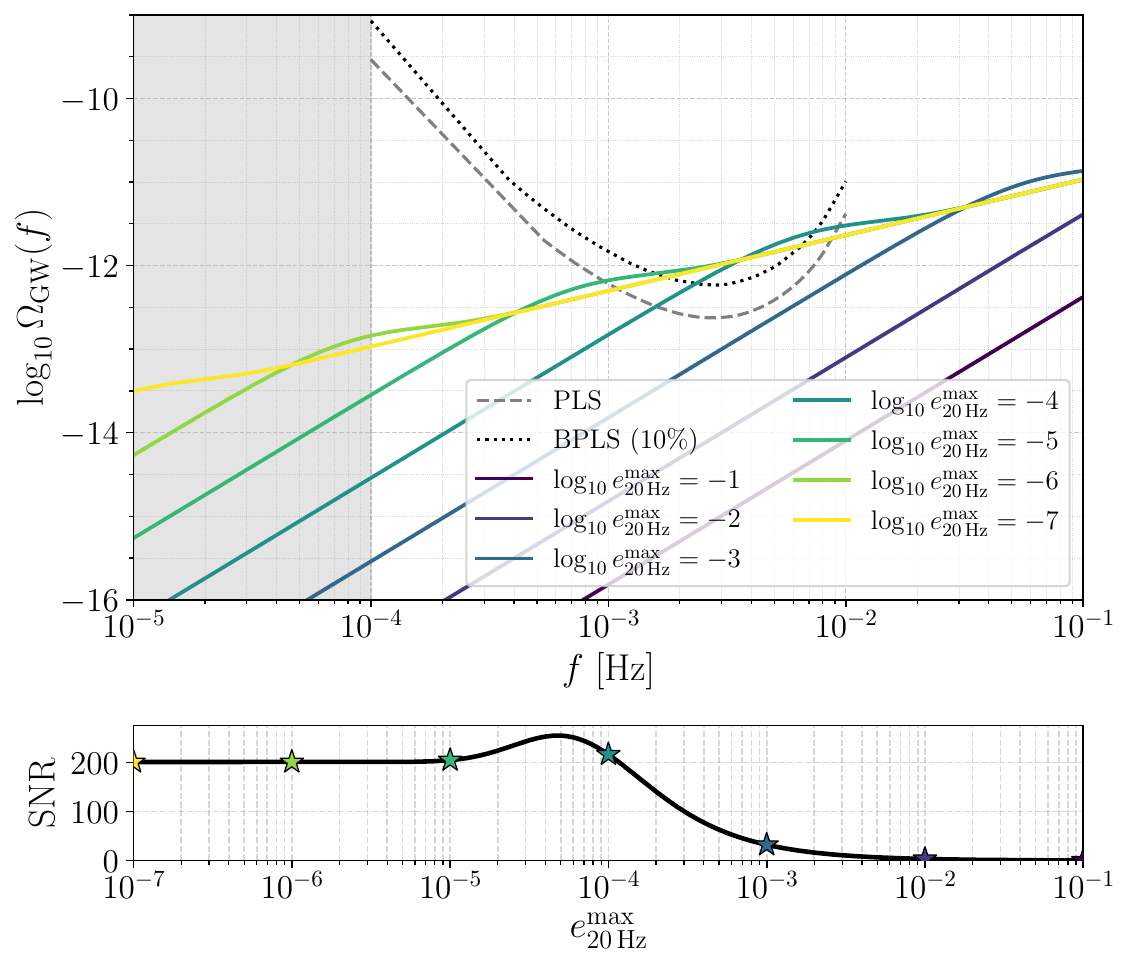}
    \caption{Top: SGWB spectra $\Omega_{\mathrm{GW}}(f)$ for sBBH populations with a uniform eccentricity distribution between $0$ and $e^{\mathrm{max}}_{20\,\mathrm{Hz}}$ at the LVK reference frequency of $20\,\mathrm{Hz}$, for $\log_{10} e^{\mathrm{max}}_{20\,\mathrm{Hz}}\in\{-1,-2,\ldots,-7\}$. For reference, we also show the LISA PLS (gray dashed) and BPLS at $10\%$ noise uncertainty (black dotted) curves; the grey shaded region marks frequencies outside the LISA sensitivity band. Bottom: LISA-band SNR as a function of $e^{\mathrm{max}}_{20\,\mathrm{Hz}}$.}
    \label{fig:induced_erefmax_gwb_snr}
\end{figure}

The resulting spectra are shown in the top panel of~\cref{fig:induced_erefmax_gwb_snr}, while the bottom panel shows the corresponding LISA-band SNR as a function of $e^{\mathrm{max}}_{20\,\mathrm{Hz}}$. 
We compute the SNR as
\begin{equation}
\mathrm{SNR}=\sqrt{T_{\rm obs} \sum_k 4 \int_0^{\infty} d f\left(\frac{S_{k, \mathrm{GW}}(f)}{S_{k, \mathrm{n}}(f)}\right)^2},
\end{equation}
where $S_{k, \rm GW}(f)$ is the PSD of the astrophysical GW stochastic signal, and $S_{k,\rm n}(f)$ is the instrumental PSD, in a given TDI channel, with $T_{\rm obs} = 4 \mathrm{yr}$.
For sufficiently small values, $e^{\mathrm{max}}_{20\,\mathrm{Hz}}\lesssim 10^{-5}$, the eccentricity in the LISA band is still negligible and the spectrum is essentially indistinguishable from the quasi-circular prediction, giving a roughly constant SNR of $\sim 200$.
As $e^{\mathrm{max}}_{20\,\mathrm{Hz}}$ increases, the backward evolution produces large eccentricities in the LISA band, redistributing power into higher harmonics and thereby strongly suppressing the SGWB at mHz frequencies.
The SNR correspondingly drops by nearly two orders of magnitude, reaching a value of $\sim 3$ already at $e^{\mathrm{max}}_{20\,\mathrm{Hz}} = 10^{-2}$, and the spectrum falls well below the LISA BPLS curve for $e^{\mathrm{max}}_{20\,\mathrm{Hz}}=10^{-1}$.

We have thus illustrated the following: a confident LISA detection of the sBBH stochastic signal, consistent with the quasi-circular expectation, would effectively rule out $e^{\mathrm{max}}_{20\,\mathrm{Hz}}\sim 0.01$ or larger, assuming for illustrative purposes that $e_{20 \, \mathrm{Hz}}$ follows a uniform distribution. 
Meanwhile, a stochastic signal that is suppressed in the LISA band would suggest relatively large eccentricities for the sBBH population in the LVK band.
One can quantify this further by performing a dedicated Bayesian analysis with  $e^{\mathrm{max}}_{20\,\mathrm{Hz}}$ as a population parameter, which we leave to future work.
For the case of the uniform distribution we considered, the illustrative bounds that we have outlined in this Section essentially depend only the mean value
of the eccentricity $\bar{e}_{20\,\mathrm{Hz}}$ across the population.
We also stress that the strength of these bounds depends sensitively on the assumed shape of the eccentricity distribution~\cite{2020ApJ...903...67M,Xuan:2024dvx,Liang:2025wfj}: a log-uniform prior, for instance, would
concentrate probability mass near $e_{\rm 20 Hz}=0$ and essentially wash out the bound on $e^{\mathrm{max}}_{20\,\mathrm{Hz}}$.
In general, we expect that one can try to constrain the mean, variance, and skewness of the $e_{20\,\mathrm{Hz}}$ distribution.

Nevertheless, the present analysis already shows that LISA can translate a non-detection of eccentricity effects in the mHz band into a meaningful constraint on the eccentricity of the sBBH population at LVK frequencies.
This is complementary to recent LVK studies showing how eccentricity non-detections in resolved sBBH events can constrain dynamical formation channels~\cite{2026arXiv260329019G}.
We stress that such a constraint would be particularly valuable for the next generation of ground-based detectors, such as the Einstein Telescope and Cosmic Explorer, which will be operating when LISA flies, and whose data analysis will benefit from independent information on the astrophysical distribution of sBBH eccentricity near merger.

\section{Conclusion}
\label{sec:conclusion}
The SGWB from sBBHs in the LISA band can probe their astrophysical formation history, with LISA being sensitive to the early inspiral where residual eccentricity may still be significant.
In such cases, the resulting stochastic background can deviate appreciably from the standard quasi-circular prediction, as eccentricity redistributes power into higher harmonics and thereby suppresses the spectrum at low frequencies relative to the circular case.
In this work, we developed an improved SGWB model containing eccentricity effects, which we applied to both an idealized Dirac-delta distribution and a thermal distribution for eccentricity.
Using a fully Bayesian framework, we then investigated how eccentricity can be measured from the LISA detection of the resulting SGWB. 

Specifically, for the Dirac-delta distribution, we found that sufficiently large injected eccentricities lead to informative posterior constraints on $e_0$. 
For $e_0 \gtrsim 0.9$ at an orbital frequency $f_{\rm orb}=10^{-4}\,\mathrm{Hz}$, Bayesian model selection strongly favors the eccentric model over the quasi-circular model.
We then applied this model to both a phenomenological mixture model and a more astrophysically motivated case of a thermal eccentricity distribution. 
For the mixture model, the evidence for eccentricity increases with the mixing fraction, indicating that even for highly eccentric binaries, a sufficiently large eccentric subpopulation is needed for its imprint to be robustly identified in the stochastic signal. 
For the thermal eccentricity distribution, the SGWB is effectively consistent with a circular model when binaries form at $f_{\rm orb} = 10^{-5}\,\mathrm{Hz}$, whereas formation at $f_{\rm orb} = 10^{-4}\,\mathrm{Hz}$ can induce appreciable systematic biases relative to the circular power law model prediction. Thus, even when eccentricity is not explicitly modeled for, 
such as in a power law model, one can identify departures from the prediction based on circular orbits.

We further studied the degeneracy between eccentricity and non-vacuum effects (due to astrophysical environment), focusing on dynamical friction as an example.
Specifically, we computed the Bayes factor between the non-vacuum model and the vacuum model, with eccentricity effects included in the latter. 
As expected, we found that including eccentricity suppresses the evidence in favor of the non-vacuum model, and only relatively dense environments with the disk gas density of $\rho \gtrsim 10^{-7} \mathrm{g} \mathrm{cm}^{-3}$ can be distinguished from the vacuum case.

Finally, we considered a sBBH population with a uniform eccentricity distribution between $0$ and a maximum value $e^{\mathrm{max}}_{20\,\mathrm{Hz}}$, defined at a ground-based detector reference GW frequency of $20\,\mathrm{Hz}$. 
We showed that the corresponding LISA SGWB is strongly suppressed already for $e^{\mathrm{max}}_{20\,\mathrm{Hz}}\sim 0.01$. As a result, a LISA detection of the sBBH SGWB consistent with the quasi-circular expectation would translate into an upper bound on the eccentricity of the sBBH population in the band of ground-based detectors, with direct implications for template modeling and data analysis in future ground-based observatories operating alongside LISA.

As a first step, our analysis adopted simplified eccentricity distributions and fixed characteristic formation frequencies. A more realistic treatment should account for the full population distribution of eccentricity and its correlation with other binary parameters, such as mass, redshift, and formation channel. 
In addition, environmental effects themselves can alter eccentricity evolution~\cite{Naoz:2012bx,Chandramouli:2021kts,Roedig:2011,Romero-Shaw:2024klf,ONeill:2024tnl}, so a fully self-consistent treatment of eccentric and non-vacuum populations remains an important direction for future work.
Another interesting direction is to explore the implications of the universal eccentricity distribution, computed recently in Ref.~\cite{Rozner:2026jtj}, on the SGWB from the sBBH in the LISA band.
Overall, these findings underscore the importance of including eccentricity in SGWB modeling in order to enable reliable astrophysical interpretation of future LISA observations.

\begin{acknowledgments}
The authors are grateful to A. Sesana, M. Dotti, D. Gerosa, A. Toubiana, F. Duque, J. R. Gair and S. Chen for valuable inputs and helpful discussion. 
R.C is supported by the China Scholarship Council (No.202406340069), the National Natural Science Foundation of
China (Grant No.12233011).
E.B. and R.S.C. acknowledge support from the European Union’s Horizon ERC Synergy Grant ``Making Sense of the Unexpected in the Gravitational-Wave Sky'' (Grant No. GWSky-101167314) and the PRIN 2022 grant ``GUVIRP - Gravity tests in the UltraViolet and InfraRed with Pulsar timing''. F.P. is supported by MPG. R.B. acknowledges financial support by the Italian Space Agency grant \emph{Phase
B2/C activity for LISA mission, Agreement n.2024-NAZ-0102/PE}. 
\end{acknowledgments}
\appendix

\section{Validity of eccentric SGWB model}\label{app:validity} 

\subsection{Alternative derivation of eccentric SGWB}\label{app:validity_reaction} 
For a single eccentric binary, there can be interference between Fourier harmonics due to radiation reaction, which for example is clearly shown in the Fourier spectrum of the waveform in Fig.~3 of Ref.~\cite{Moore:2018kvz}.
Qualitatively, such interference effects happens when the evolution of a given harmonic exceeds the frequency bin width and the initial spacing between harmonics.
When a binary is widely separated, the orbital back-reaction is entirely negligible, resulting in a discrete Dirac-delta spectrum.
At closer separations, there will be appreciable back-reaction that results in interference between the harmonics. 
We also know from Phinney's theorem~\cite{Phinney:2001di} that there is a general equivalence between the waveform polarizations and the energy spectrum.
Can the interference terms that appear in the waveform polarizations then contribute to the energy spectrum?
We show below that these interference terms in fact vanish when averaging over all orientations of the binary, and obtain the same expression for the energy spectrum given in~\cref{eqn:energy_spectrum}.

To leading PN order, the frequency-domain GW polarizations for an eccentric binary computed using the stationary-phase approximation is given by~\cite{Yunes:2009yz,Moore:2018kvz,Chandramouli:2021kts}
\begin{align}\label{eqn:GW_SPA}
\tilde{h}_{+,\times} &= -\dfrac{\eta m}{ 2 D_L} \sum_{j=1}^{\infty} \dfrac{[2 \pi m f_{\rm orb} (t_j^*)]^{2/3} }{ \sqrt{j \dot{f}_{\rm orb}(t_j^*)}} \nonumber \\
&\times \left [ C^{(j)}_{+,\times}(t_j^*) + i  S^{(j)}_{+,\times}(t_j^*)  \right] e^{-i \psi_j(t_j^*)},
\end{align}
where $t_j^*$ is the stationary point where $f_{\rm orb}(t_j^*) = f_r(t_j^*) / j = f (1+z)/j$, and $\psi_j (t_j^*)$ is the Fourier GW phase evaluated at the stationary time. 
The Fourier coefficients of the polarizations $C_{+,\times}^{(j)}$ and $S_{+,\times}^{(j)}$ are given by
\begin{subequations}\label{eqn:Fourier_polarizations}
\begin{align}
C_{+}^{(j)} & =\left[2 s_\iota^2 J_j(j e)+\frac{2}{e^2}\left(1+c_\iota^2\right) c_{2 \beta}\left(\left(e^2-2\right) J_j(j e)\right.\right. \nonumber \\
& \left.\left.+j e\left(1-e^2\right)\left(J_{j-1}(j e)-J_{j+1}(j e)\right)\right)\right], \\
S_{+}^{(j)} & =-\frac{2}{e^2} \sqrt{1-e^2}\left(1+c_\iota^2\right) s_{2 \beta}\left[-2\left(1-e^2\right) j J_j(j e)\right. \nonumber \\
& \left.+e\left(J_{j-1}(j e)-J_{j+1}(j e)\right)\right], \\
C_{\times}^{(j)} & =-\frac{2}{e^2} c_\iota s_{2 \beta}\left[2\left(e^2-2\right) J_j(j e)\right. \nonumber\\
& \left.+2 j e\left(1-e^2\right)\left(J_{j-1}(j e)-J_{j+1}(j e)\right)\right] \\
S_{\times}^{(j)} & =-\frac{4}{e^2} \sqrt{1-e^2} c_\iota c_{2 \beta}\left[-2\left(1-e^2\right) j J_j(j e)\right. \nonumber \\
& \left.+e\left(J_{j-1}(j e)-J_{j+1}(j e)\right)\right],
\end{align}
\end{subequations}
where we used the shorthand $c_\iota = \cos \iota$, $s_{\iota} = \sin \iota$, $c_{2 \beta} = \cos 2\beta$, $s_{2\beta} = \sin 2\beta$, with $\iota$ being the inclination angle and $\beta$ being the polarization angle. Note that we specifically used the expressions as given in Ref.~\cite{Chandramouli:2021kts} that had corrected typos in previous literature.
The Fourier GW phase of the $j$-th harmonic is of the form,
\begin{align}
\psi_j (t_j^*) = 2 \pi t_c f_r ( t_j^*) - j \ell_c + \Phi_j (e(t_j^*)),
\end{align}
where $t_c$ and $\ell_c$ are constants of integration that specify a reference value of the binary's mean anomaly $\ell$, and $\Phi_j(e(t_j^*))$ is the non-trivial GW phasing that depends on $f_r$ implicitly through $e(t_j^*)$.
It is important to recognize that $e(t_j^*) = e_j$ as defined in~\cref{eqn:eccn}.
The expression for $\Phi_j$, valid for arbitrary eccentricity, is rather cumbersome and is given in terms of hypergeometric functions in Ref.~\cite{Moore:2018kvz}, but it is not needed for our calculation.
For completeness, note that in the circular limit, the contribution is from $j=2$ and the GW dephasing is given by $\Phi_2 (t_2^*)  = 3/128 (2 \pi \mathcal{M} f_{\rm orb}(t_2^*))^{-5/3}$, which is found in textbooks, such as Ref.~\cite{Maggiore:1999vm}.

From Phinney's theorem~\cite{Phinney:2001di}, we have that
\begin{align}\label{eqn:phinney_theorem}
\dfrac{dE_{\rm GW}}{df_r} &=  2\pi^2 \dfrac{D_L^2}{(1+z)^2} f^2 \nonumber \\
& \times  \left.\left ( \langle \left \vert  \tilde{h}_{+} (f_r) \right \vert^2 \rangle + \langle \left \vert  \tilde{h}_{\times} (f_r) \right \vert^2 \rangle \right) \right \vert_{f_r = f (1+z)},
\end{align}
where $\langle X \rangle$ represents an average over the binary's orbital orientation,
\begin{align}
\langle X \rangle = \dfrac{1}{8 \pi^2} \int_0^{\pi }\mathrm{d} \iota \sin \iota  \int_0^{2\pi } \mathrm{d} \beta \int_0^{2\pi } \mathrm{d} \ell_c.
\end{align}
We first compute $ \left \vert  \tilde{h}_{+,\times} (f_r) \right \vert^2$ from~\cref{eqn:GW_SPA},
\begin{widetext}
\begin{align}\label{eqn:polarization_square}
 \left \vert  \tilde{h}_{+,\times} (f_r) \right \vert^2 & = \dfrac{\eta^2 m^2}{ 4 D_L^2} \sum_{j=k=1}^{\infty}  \dfrac{[2 \pi m f_{\rm orb} (t_j^*)]^{4/3}}{j \dot{f}_{\rm orb}(t_j^*)} \left[ (C_{+,\times}^{(j)} (t_j^*))^2 + (S_{+,\times}^{(j)} (t_j^*))^2 \right]  \nonumber \\
 & + \dfrac{\eta^2 m^2}{ 4 D_L^2} \sum_{j \neq k} \sum_{k=1}^{\infty}  \dfrac{[4 \pi^2 m^2 f_{\rm orb} (t_j^*) f_{\rm orb} (t_k^*)]^{2/3}}{\sqrt{j k \dot{f}_{\rm orb}(t_j^*) \dot{f}_{\rm orb}(t_k^*)}} e^{-i (\Psi_j (t_j^*) - \Psi_k (t_k^*))} \nonumber \\
  \times & \left[ C_{+,\times}^{(j)} (t_j^*) C_{+,\times}^{(k)} (t_k^*) + S_{+,\times}^{(j)} (t_j^*) S_{+,\times}^{(k)} (t_k^*) - i C_{+,\times}^{(j)} (t_j^*) S_{+,\times}^{(k)} (t_k^*) + i C_{+,\times}^{(k)} (t_k^*) S_{+,\times}^{(j)} (t_j^*)\right],
\end{align}
\end{widetext}
where we have split the sum into the diagonal and off-diagonal contributions.
The second term, which is the off-diagonal contribution, is the source of the interference effect for a given binary as it depends on the phase difference $\Psi_j (t_j^*) - \Psi_k (t_k^*)$.
This phase difference depends on $(j - k) \ell_c$, and therefore the off-diagonal contribution has an overall multiplicative factor of $\exp \left ( i (j - k) \ell_c \right)$. 
Thus, when performing the average over $\ell_c$, this off-diagonal contribution will vanish.
In other words, for randomly oriented phases, there is no interference effect between the Fourier harmonics in the SGWB (assuming the continuum limit).

To compute the diagonal contributions, we make use of the following:
\begin{widetext}
\begin{subequations}
\begin{align}
\left\langle \big(C_{+}^{(j)}(t_j^*)\big)^2 \right\rangle
&= \frac{8}{15e_j^{4}}\Bigg[
21e_j^{2}(1-e_j^{2})^{2}j^{2}J_{j-1}(j e_j)^{2} 
+28e_j(1-e_j^{2})j\big(2+j-e_j^{2}(1+j)\big)J_{j-1}(j e_j)J_{j}(j e_j)\nonumber\\
&\qquad+\big(28+56j-28e_j^{2}(1+3j)+e_j^{4}(11+28j)\big)J_{j}(j e_j)^{2} 
+7e_j^{2}(1-e_j^{2})^{2}j^{2}J_{j+1}(j e_j)^{2}
\Bigg],\\
\left\langle \big(S_{+}^{(j)}(t_j^*)\big)^2 \right\rangle
&= -\frac{224(1-e_j^{2})\left(e_j\,J_{j-1}(j e_j)-\big(1+(1-e_j^{2})j\big)J_{j}(j e_j)\right)^{2}}{15e_j^{4}}, \\
\left\langle \big(C_{\times}^{(j)}(t_j^*)\big)^2 \right\rangle
&= \frac{8}{3e_j^{4}}
\left(
2e_j(1-e_j^{2})j\,J_{j-1}(j e_j)
+\big(2-e_j^{2}+2(1-e_j^{2})j\big)J_{j}(j e_j)
\right)^{2}, \\
\left\langle \big(S_{\times}^{(j)}(t_j^*)\big)^2 \right\rangle
&= \frac{32(1-e_j^{2})}{3e_j^{4}}
\left(
e_j\,J_{j-1}(j e_j)
-\big(1+(1-e_j^{2})j\big)J_{j}(j e_j)
\right)^{2}.
\end{align} \label{eqn:Fourier_coeff_square_avg}
\end{subequations}
\end{widetext}
We define the $j$-th term of the energy spectrum as $(dE_{\rm GW}/ df_r)_j$, so that the total spectrum given by~\cref{eqn:phinney_theorem} reads $dE_{\rm GW}/ df_r = \sum_{j=1}^{\infty} (dE_{\rm GW}/ df_r)_j $.
With~\cref{eqn:polarization_square,eqn:Fourier_coeff_square_avg} in hand, $(dE_{\rm GW}/ df_r)_j$ becomes
\begin{widetext}
\begin{align}
\left( \dfrac{dE_{\rm GW}}{df_r} \right)_j &= \frac{(1-e_j^{2})^{7/2}\,\mathcal{M}_c^{5/3}\,j^{4/3}(2\pi)^{2/3}}
{3e_j^{4}\!\left(96+292e_j^{2}+37e_j^{4}\right) f^{1/3}(1+z)^{1/3}}
\Bigg[
e_j^{2}(1-e_j^{2})\!\left(48+41(1-e_j^{2})j^{2}\right) J_{j-1}(j e_j)^{2}  \nonumber \\
&-4e_j(1-e_j^{2})\!\left(24+12(4-3e_j^{2})j+17(1-e_j^{2})j^{2}\right)
J_{j-1}(j e_j)J_{j}(j e_j) \\
&-4\!\left(
-24-48j-17j^{2}
+12e_j^{6}j^{2}
-e_j^{4}(4+36j+41j^{2})
+e_j^{2}(24+84j+46j^{2})
\right) J_{j}(j e_j)^{2} \nonumber \\
&+7e_j^{2}(1-e_j^{2})^{2}j^{2}J_{j+1}(j e_j)^{2} \nonumber
\Bigg].
\end{align}
\end{widetext}
Using the recurrence relations for the Bessel functions~\cite{AbramowitzStegun,WatsonBessel} given by
\begin{subequations}
\begin{align}
J_{j-2}(x) &= \frac{2(j-1)}{x} J_{j-1}(x) - J_j(x), \\
J_{j+2}(x) &= \frac{2(j+1)}{x} J_{j+1}(x) - J_j(x), \\
J_{j+1}(x) &= \frac{2j}{x} J_j(x) - J_{j-1}(x),
\end{align}
\end{subequations}
and the expression for $\dot{f}_{\rm orb}$ from~\cref{eqn:dforbdt}, we simplify $(dE_{\rm GW}/ df_r)_j$ and obtain the total energy spectrum as
\begin{equation}
\begin{aligned}
\frac{d E_{\mathrm{GW}}}{d f_{r}} &= \frac{\mathcal{M}^{5/3} \pi^{2/3}}{3f^{1/3}(1+z)^{1/3}} \sum_{j=1}^{\infty}\frac{g(j,e_j)}{F(e_j)(j/2)^{2/3}},\\
\end{aligned}
\end{equation}
which reproduces~\cref{eqn:energy_spectrum} and completes the alternative derivation of the energy spectrum.

\subsection{Validity of neglecting higher PN contributions}\label{app:validity_PN} 
In this work, we have neglected PN corrections to the modeling of the SGWB.
Typically, the PN corrections get enhanced as eccentricity increases, and thus one needs to determine at what eccentricity the leading PN approximation breaks down or is inaccurate.
From Ref.~\cite{Arun:2007sg}, the ratio of the instantaneous orbit-averaged energy fluxes between the 1PN and the 0PN contributions are given by
\begin{widetext}
\begin{align}
\dfrac{\mathcal{F}_{1 \rm PN,inst}}{\mathcal{F}_{0 \rm PN, inst}} &= \dfrac{x}{1-e_t^2} \left[ \dfrac{ -\left(\dfrac{1247}{336} + \dfrac{35}{12} \eta  \right)+ \left(  \dfrac{10475}{672} - \dfrac{1081}{36} \eta \right) e_t^2 + \left(  \dfrac{10043}{384} - \dfrac{311}{12} \eta \right) e_t^4 + \left(  \dfrac{2179}{1792} - \dfrac{851}{576} \eta \right) e_t^6 }{1 + \dfrac{73}{24} e_t^2 + \dfrac{37}{96}e_t^4 }\right], \label{eqn:PN_fluxes_any_ecc}
\end{align}
\end{widetext}
where $e_t$ is the \emph{time eccentricity} associated with the PN accurate quasi-Keplerian parametrization of the osculating orbit~\cite{Damour_Deruelle:19851} and  $x = (2\pi m f_{\rm orb})^{2/3}$ is the gauge-invariant PN expansion parameter~\cite{Arun:2007sg}. 
Note that the Keplerian eccentricity $e$ used in our work can be related to the time eccentricity $e_t$, using the expressions for each quantity written in terms of the gauge-invariant energy and angular momentum~\cite{Damour_Deruelle:19851} (see Ref.~\cite{Gamboa:2024imd} for an explicit demonstration up to 3PN order).
In the small eccentricity asymptotic limit $e_t \ll 1$,~\cref{eqn:PN_fluxes_any_ecc} becomes
\begin{align}
\left (\dfrac{\mathcal{F}_{1 \rm PN,inst}}{\mathcal{F}_{0 \rm PN, inst}} \right)_{e_t \ll 1} \sim - x \left(\dfrac{1247 + 980 \eta}{336} \right), \label{eqn:PN_fluxes_low_ecc}
\end{align}
and in the high eccentricity asymptotic limit $1-e_t \ll 1$,~\cref{eqn:PN_fluxes_any_ecc} instead becomes
\begin{align}
\left (\dfrac{\mathcal{F}_{1 \rm PN,inst}}{\mathcal{F}_{0 \rm PN, inst}} \right)_{1-e_t \ll 1} \sim \dfrac{x}{1-e_t^2} \left(\dfrac{30141-46340 \eta}{3400} \right). \label{eqn:PN_fluxes_high_ecc}
\end{align}
Thus the smallness of the PN expansion is governed by $x$ for small eccentricity, and $x/(1-e_t^2)$ for large eccentricity, with a weak $\mathcal{O}(1)$ dependence on the symmetric mass ratio.

Using~\cref{eqn:PN_fluxes_low_ecc}, for typical sBBHs with comparable  masses $m_1 \sim m_2 $, we find that the ratio of the fluxes is $\lesssim 1/10$ when
\begin{align}
f_{\rm orb} \lesssim 2.1 \mathrm{Hz} \times \left ( \dfrac{50 M_{\odot}}{m} \right), \label{eqn:validity_forb}
\end{align}
and thus we can neglect PN contributions in the LISA band when computing the SGWB for small eccentricities.
Meanwhile, using~\cref{eqn:PN_fluxes_high_ecc} for high eccentricities (and again comparable masses), requiring that the ratio of fluxes remain $\lesssim 1/10$ results in
\begin{align}
1-e_t \gtrsim 7.9 \times 10^{-4} \times \left ( \dfrac{m}{50 M_{\odot}} \right) ^{2/3} \left ( \dfrac{f_{\rm orb}}{10^{-4} \mathrm{Hz}} \right) ^{2/3}. \label{eqn:validity_ecc}
\end{align}
This means that our eccentric SGWB model will remain valid for $e_0 \lesssim 0.99921$ with $f_{\rm{orb},0} = 10^{-4} \rm{Hz}$. 
When using~\cref{eqn:validity_forb} or~\cref{eqn:validity_ecc}, we caution that these should be used in the respective asymptotic limits and only provide a Fermi estimate of the relative importance of the 1PN contribution.
As a more generic case with $e_0 = 0.9$ at $f_{\rm{orb},0} = 10^{-4} \rm{Hz}$, the ratio of fluxes is $\sim 10^{-2}$ for $m_1 \sim m_2 \sim 25 M_{\odot}$.
Thus, for the goals of this work, we can neglect PN corrections in computing the SGWB in the LISA band.
\section{Additional details on Bayesian data analysis}\label{app:additional}

\begin{figure}[!t]
    \centering
    \includegraphics[width=\columnwidth]{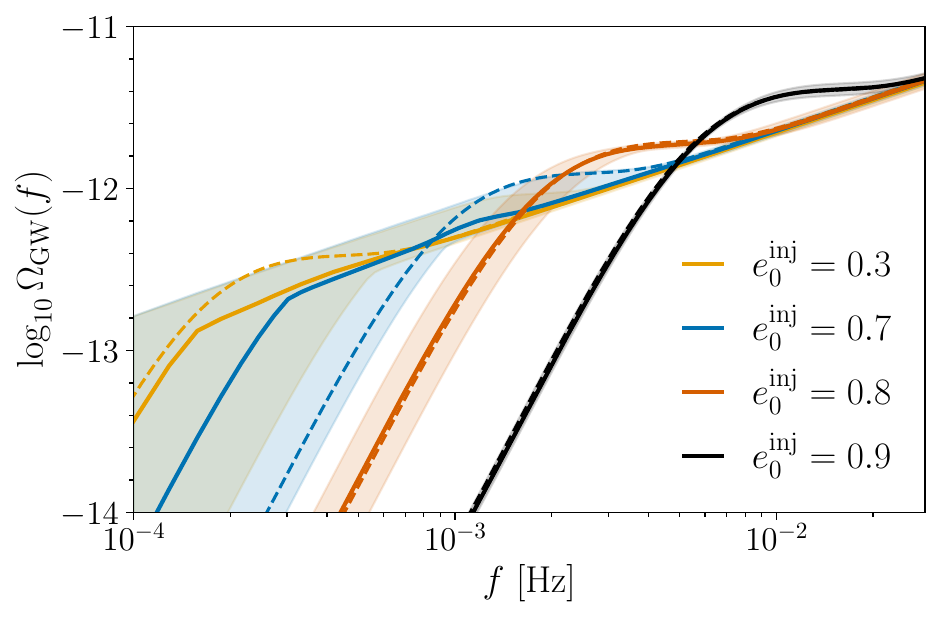}
    \caption{Posterior predictives of eccentricity model compared against the injected signals (dashed lines) for $e_{0}^{\rm{inj}} = \{ 0.3, 0.7, 0.8, 0.9\}$.
    For each case, the shaded region corresponds to the 90\% credible interval, with solid line corresponding to the respective median.
    }
    \label{fig:ecc_template_reconstruction}
\end{figure}

The marginalized posteriors shown in~\cref{fig:ecc_template_recovery} can be further understood in terms of the posterior predictives.
Specifically, we compute the eccentric model from the posterior samples at each frequency, and obtain the corresponding median and $90\%$ credible interval.
In~\cref{fig:ecc_template_reconstruction}, we show the injected signals (dashed lines) together with the posterior predictives for the cases with $e_0^{\rm inj} = \{0.3,0.7,0.8,0.9 \}$, with the median represented by solid lines and the shaded region corresponding to the $90\%$ credible intervals.
For high injected $e_{0}\geq0.8$, the median posterior predictive is accurate compared to the injected signal, consistent with the marginalized posteriors (cf.~\cref{fig:ecc_template_recovery}), which show a clear peak for $e_0$ that is close to the injected value.
In contrast, for lower eccentricities, there is mismatch between the median posterior predictive and the injected signal, but the differences are within the $90\%$ credible intervals, which is also consistent with obtaining a rather uninformative marginalized posterior for $e_0$ (cf.~\cref{fig:ecc_template_recovery}).

\begin{figure}[!t]
    \centering
    \includegraphics[width=\columnwidth]{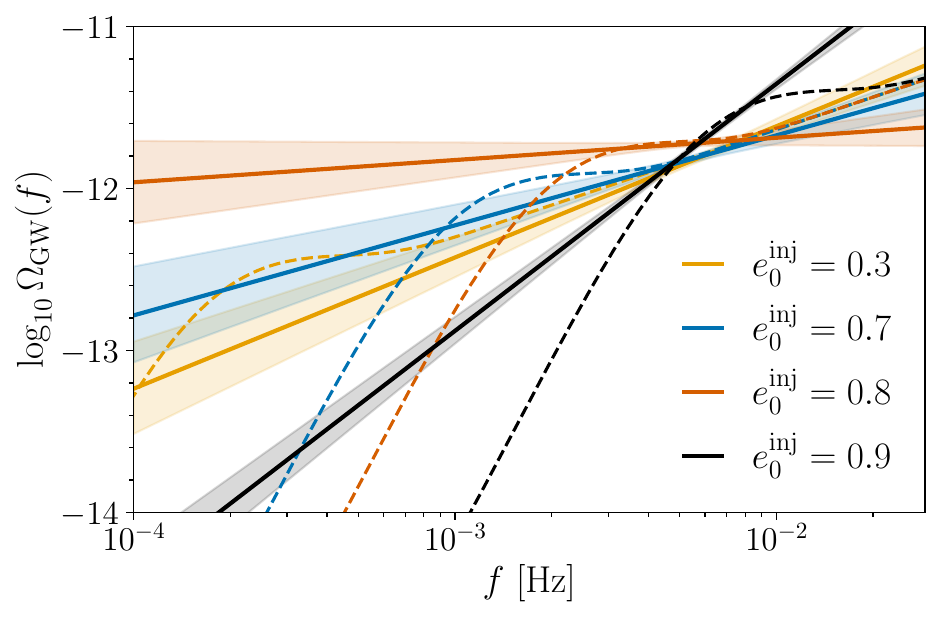}
    \caption{Posterior predictives of the power law model compared against the injected signals (dashed lines) for $e_{0}^{\rm{inj}} = \{ 0.3, 0.7, 0.8, 0.9\}$.
    For each case, the shaded region corresponds to the 90\% credible interval, with solid line corresponding to the respective median.
    }
    \label{fig:PL_template_reconstruction}
\end{figure}

In~\cref{tab:posterior_statistical_uncertainties_delta_e0}, we found a non-trivial trend for the Bayes factors between the eccentric and circular models.
To better understand this behavior, we now turn to~\cref{fig:PL_template_reconstruction} which shows the posterior predictives obtained with the circular (power law) model, given the injected eccentric spectra.
We find that the posterior predictives for each case have different spectral slopes, as $\gamma$ (and $A_{\rm vac}$) is biased relative to its true circular value.
For eccentricities $e_0^{\rm inj} = \{ 0.3 , 0.7 \}$, the higher frequency content (in the range $\sim 10^{-3}$---$10^{-2}$Hz), where LISA is sensitive, is well captured by the median posterior predictive of the circular model, resulting in only marginal preference for the eccentric model.
For $e_0^{\rm inj} = 0.8$, since the turning point shifts to sensitive regime of the LISA band, the posterior predictive of circular (power law) model captures the injected signal very well, especially close to the peak.
In fact, the spectral slope of the median posterior predictive very closely matches the spectral slope at the peak of the injected spectrum.
However, the higher frequency content is not well captured, with the injection lying outside the $90\%$ credible interval for $f \gtrsim 10^{-2}$Hz.
Thus, there is even less evidence in favor of the eccentric model for $e_0^{\rm inj} = 0.8$, when compared to cases with $e_0^{\rm inj} \leq 0.7$, where the peak of the spectrum is not well captured by the circular model.
For a larger eccentricity of $e_0^{\rm inj} = 0.9$, there is clear mismatch between the posterior predictive of the circular model and the injected signal across the sensitive frequency band, resulting in strong evidence in favor of the eccentric model.

We point out that the log Bayes factors between the eccentric and circular (power law) models $\log \mathcal{B}^{\rm ecc}_{\rm cir}$ do not correspond to the Savage-Dickey~\cite{Chatziioannou:2014bma} estimate.
This is because when recovering with the circular (power law) model, we sample on $\gamma$, and instead keep $\gamma = 2/3$ when recovering with the eccentric model.
In other words, one should really view this as a model comparison between the eccentric model and the power law model.
For a strict model selection between the eccentric and circular hypotheses, one has to keep $\gamma = 2/3$ [as predicted by~\cref{eqn:hc_circ_pop}] when recovering with the circular model.
With $\gamma = 2/3$ fixed, the models are nested, and one can equivalently compute the Bayes factor using the Savage-Dickey ratio. 
However, as there can be errors in the Savage-Dickey estimate due to poor sampling (and binning) at $e_0 = 0$~\cite{Chatziioannou:2014bma}, we instead estimate $\log  \mathcal{B}^{\rm ecc}_{\rm cir}$ from the evidences obtained from a separate set of parameter estimation recoveries. 
We find
$\log \mathcal{B}^{\rm ecc}_{\rm cir} \approx \{-0.6515,\allowbreak -0.4814,\allowbreak -0.5612,\allowbreak -0.6105,\allowbreak -0.6074,\allowbreak -0.5445,\allowbreak -0.4242,\allowbreak 6.6178,\allowbreak 40.68\}$ for $e_0^{\rm inj} = \{0.1,0.2,0.3,0.4,0.5,0.6,0.7,0.8,0.9\}$,
and thus, the Bayes factors clearly increase with $e_0^{\rm inj}$ when keeping $\gamma = 2/3$ fixed, in contrast to the more non-trivial trend we found when sampling on $\gamma$.

\bibliographystyle{apsrev4-2}
\bibliography{ref.bib}
\end{document}